\documentclass[journal]{IEEEtran}
\usepackage{cite}

\ifCLASSINFOpdf
  \usepackage[pdftex]{graphicx}
  \graphicspath{{./figs/}}
  \DeclareGraphicsExtensions{.pdf,.jpg,.png,.eps}
\else
\fi
\usepackage{xcolor}

\usepackage{amsmath}
\usepackage{amssymb}

\def\Gpqn{{\mathord{\buildrel{\lower3pt\hbox{$\scriptscriptstyle\leftrightarrow$}} 
\over G}_{\raisemath{1pt}{pq}}^{\raisemath{-4pt}{\nu}} }}
\def\tensor#1{\mathord{\buildrel{\lower3pt\hbox{$\scriptscriptstyle\leftrightarrow$}} 
\over{#1}}}
\def\cup#1{^{\raisemath{-4pt}{#1}}}
\def\mm#1#2{\raisemath{#1 pt}{#2}}
\mathchardef\mhyphen="2D
\def\rar#1{\renewcommand{\arraystretch}{#1}}

\makeatletter
\newcommand{\raisemath}[1]{\mathpalette{\raisem@th{#1}}}
\newcommand{\raisem@th}[3]{\raisebox{#1}{$#2#3$}}
\makeatother

\usepackage[nomessages]{fp}
\usepackage[percent]{overpic}
\usepackage{pict2e}
\usepackage{tabularx}
\usepackage{cuted}
\usepackage{subcaption}
\usepackage{tikz}
\usetikzlibrary{angles}
\usepackage{mathtools}
\usepackage{multirow}

\begin{document}

\title{Efficient Computation of Spatially-Discrete Traveling-Wave Modulated Structures}

\author{
Cody Scarborough, \IEEEmembership{Member, IEEE, }Zhanni Wu, \IEEEmembership{Member, IEEE,} and 
Anthony Grbic, \IEEEmembership{Fellow, IEEE }%
\thanks{C. Scarborough is with the Department of Electrical Engineering and Computer Science, University of Michigan, 1301 Beal Avenue, Ann Arbor, MI 48109-2122, USA (e-mail: codyscar@umich.edu).}
\thanks{Z. Wu is with the Department of Electrical Engineering and Computer Science, University of Michigan, 1301 Beal Avenue, Ann Arbor, MI 48109-2122, USA (e-mail: zhanni@umich.edu).}
\thanks{A. Grbic is with the Department of Electrical Engineering and Computer Science, University of Michigan, 1301 Beal Avenue, Ann Arbor, MI 48109-2122, USA (e-mail: agrbic@umich.edu).}}

%
%


\maketitle

\begin{abstract}
Traveling-wave modulation is a form of space-time modulation which has been shown to enable unique electromagnetic phenomena such as non-reciprocity, beam-steering, frequency conversion, and amplification. In practice, traveling-wave modulation is achieved by applying a staggered time-modulation signal to a spatially-discrete array of unit cells. Therefore, the capability to accurately simulate spatially-discrete traveling-wave modulated structures is critical to design. However, simulating these structures is challenging due to the complex space-time dependence of the constituent unit cells. In this paper, a field relation (referred to as the interpath relation) is derived for spatially-discrete traveling-wave modulated structures. The interpath relation reveals that the field within a single time-modulated unit cell (rather than an entire spatial period) is sufficient to determine the field solution throughout space. It will be shown that the interpath relation can be incorporated into existing periodic method of moments solvers simply by modifying the source basis functions. As a result, the computational domain is reduced from an entire spatial period to a single time-modulated unit cell, dramatically reducing the number of unknowns. In the context of traveling-wave modulation, this enables researchers to efficiently simulate both complex structures with patterned unit cells in addition to continuous structures with infinitesimal unit cells.
\end{abstract}


%
\IEEEpeerreviewmaketitle

\section{Introduction}

\FPset\rc{-12.5}
\FPset\rL{-9}
\FPset\rS{63}
\begin{figure}
\begin{tikzpicture}
	\definecolor{myred}{rgb}{.6,0,0};
	\definecolor{myblue}{rgb}{0,0,.6};
	\definecolor{mylightblue}{rgb}{.4,.7,.9};
	\definecolor{mygreen}{rgb}{0,.4,0};
	\definecolor{myyellow}{rgb}{1,.98,.1};
	\definecolor{mytan}{rgb}{1,.83,.66};
	\definecolor{myorg}{rgb}{1,.6,.2};
    \node[anchor=south west,inner sep=0](image)at(0,0,0){\includegraphics[clip,trim=0cm 3cm 0cm 3cm,width=.95\columnwidth]{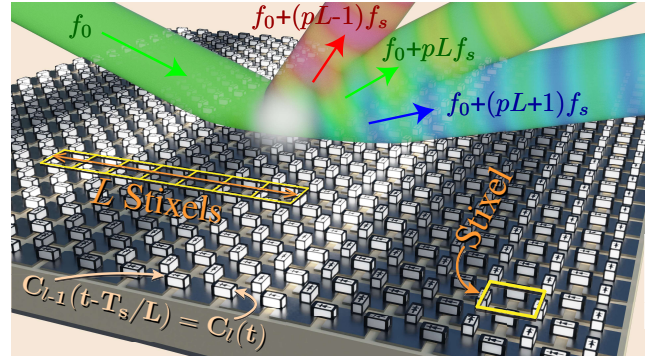}};
    \begin{scope}[x={(image.south east)},y={(image.north west)},font=\fontsize{10}{8}]
        
        \foreach \x in {0,1}{
		\node[mygreen] (c) at  (.11,.93) {  $f_0$ };
		\node[mygreen] (c) at  (.66,.865) {  $f_0$+$pLf_s$ };
		\node[myred] (c) at  (.488,.955) {  $f_0$+$(pL \mhyphen 1)f_s$ };
		\node[myblue] (c) at  (.8,.71) {  $f_0$+$(pL$+$1)f_s$ };
        }
        \begin{scope}[shift={(0.21,.125)}];
			\node[black,rotate=\rc] (c) at (.002,-.006) {$\mathbf{C}_{l\mathbf{ \mhyphen 1}}\mathbf{\left(t \mhyphen T_s/L\right)} =\mathbf{C}_l\mathbf{\left(t\right)}$};
			\node[mytan,rotate=\rc] (c) at  (0,0) {$\mathbf{C}_{l\mathbf{ \mhyphen 1}}\mathbf{\left(t \mhyphen T_s/L\right)} = \mathbf{C}_l\mathbf{\left(t\right)}$};	
		\end{scope};
        \begin{scope}[shift={(.23,.44)}];
			\node[black,rotate=\rL] (c) at (.002,-.005) {\Large $L$ Stixels};
			\node[myorg,rotate=\rL] (c) at  (0,0) {\Large $L$ Stixels};	
		\end{scope};
        \begin{scope}[shift={(.745,.42)}];
			\node[black,rotate=\rS] (c) at (.002,-.005) {\Large Stixel};
			\node[myorg,rotate=\rS] (c) at  (0,0) {\Large Stixel};	
		\end{scope};
	\end{scope};
\end{tikzpicture}
\caption{Illustration of a spatially-discrete traveling-wave modulated structure \cite{zca}.
Each spatial period is made up of $L$ ``stixels'' (indivisible unit cells).
In this example, the structure consists of metallic patches interconnected by varactor diodes placed above a grounded dielectric substrate. The varactor diodes are modulated in the form of spatially-discrete traveling-wave. That is, the varactor capacitance $C_l(t)$ within stixel $l$ is delayed in time with respect to the previous stixel $l-1$.
When the spatial period is sufficiently small and the modulation frequency $f_s=1/T_s$ is low, frequencies separated by $pLf_s$ ($p \in \mathbb{Z}$) all propagate in the same direction. }
\label{fig:fancy}
\end{figure}

\IEEEPARstart{S}{pace-time} modulation has attracted renewed interest within the field of electrodynamics. Progress in the availability/performance of tunable components and materials has drawn researchers to examine time and space-time variation as a means to achieve frequency conversion \cite{zhanni,zca,caloz,debjit,oli}, amplification \cite{tien1958traveling, tien1958parametric, heffner1958gain} and non-reciprocity \cite{harish,nonrecip, hyper, optical_leakywave}. In the late 1950s traveling-wave ferromagnetic amplifiers were explored by P. Tien and H. Suhl \cite{tien1958traveling}. In \cite{tien1958traveling}, it was shown that  amplification and frequency conversion can be simultaneously achieved by applying a traveling-wave modulation of permeability to a coupled-line system. Leading into the 1960s, E. Cassedy and A. Oliner examined the dispersion relations that arose from a medium whose permittivity is modulated by a traveling-wave \cite{oli}. Similar to coupled-mode theory applied to multiconductor transmission lines \cite{scott}, the stop bands of the modulated structure were attributed to co-directional or contra-directional coupling.
However, rather than coupling between transmission lines, the energy couples between frequency harmonics \cite{oli}.

As the performance of tunable elements such as varactor diodes, electro/acousto/magneto-optic media, phase-change materials, and 2D materials has improved, researchers have begun implementing and expanding the theory of space-time modulated media \cite{shaltout2019spatiotemporal}.
This has led to the study of space-time modulated metasurfaces \cite{zhanni,zca,caloz,nonrecip,hyper,prog}.
In these structures, tunable components are arranged or patterned onto a surface or a stack of surfaces.
In particular, significant attention has been paid to the capabilities of traveling-wave modulated metasurfaces \cite{zca,hyper,nonrecip,stdiff_grat,st_metasurf_alu,fdtd,eucap,cody_meta} such as the structure depicted in Fig. \ref{fig:fancy}.
In the continuous limit (i.e. the modulation is a continuous function of space and time), the behavior of traveling-wave modulated metasurfaces can often be predicted analytically via the Lorentz transform \cite{lorentz1937electromagnetic}.
However, in practice, traveling-wave modulation is often achieved via staggered modulation of an array of discrete unit cells. We will refer to such a configuration as spatially-discrete traveling-wave modulation (SD-TWM).
In this configuration, the individual time-modulated unit cells are referred to as stixels \cite{stixel}: space-time pixels.
The discrepancy in the response between the continuous and discrete model has been generally overlooked in design. The biasing networks and fine features within the unit cells are typically only considered in a full-wave simulation just before fabrication.

The design of traveling-wave modulated metasurfaces has disrupted the traditional workflow of RF engineers and scientists. Typically, the theoretical performance of a physical device or idealized structure can be optimized and validated using a full-wave solver or circuit simulator. A large body of work has gone into the development of numerical methods, and commercial solvers have become a critical ally to pencil and paper.
However, SD-TWM structures often require complicated biasing networks and extreme temporal variation.
From a computational standpoint, this is problematic since most commercial solvers are not optimized to solve problems which are varying in both space and time.
While modern numerical techniques (such as finite-difference time-domain method and harmonic-balance) can be used to solve space-time-dependent problems \cite{fdtd,caloz,stdiff_grat,st_metasurf_alu}, the computational cost to maintain sufficient numerical accuracy can be prohibitive. Further, physical patterned unit cells cannot be easily included in the techniques presented in \cite{caloz,st_metasurf_alu}.

In this paper, we take advantage of the space-time symmetry of SD-TWM structures to dramatically reduce the computational cost of a full-wave simulation.
A boundary condition (referred to as the interpath relation) between the stixels of a SD-TWM structure is derived and incorporated into a method of moments (MoM) solver.
In Section \ref{sec:relation}, a derivation of the interpath relation is provided. In Section \ref{sec:mom}, the interpath relation is applied to a MoM formulation for the problem of a SD-TWM sheet capacitance over a grounded dielectric. The numerical results of this implementation are reported in Section \ref{sec:results}. The conclusion is presented in Section \ref{sec:conclusion}.

%

\section{The Interpath Relation for Spatially-Discrete Traveling-Wave Modulation}\label{sec:relation}
The staggered modulation scheme used to achieve SD-TWM is reminiscent of N-path circuit networks.
An N-path network contains a set of linear, periodically time-varying (LPTV) systems (paths) connected to an input and output in parallel \cite{cody}. The defining characteristic of an N-path network is that the time-variation of each path is delayed with respect to the previous path. This means that if a time-varying component (say a time-varying capacitor) on the first path has time dependence $C_0(t)$, then the time dependence on path $l$ is $C_0(t-lT_s/L)$, where $T_s$ is the modulation period and $L$ is the number of paths in parallel.

In \cite{cody}, a relation was derived between the voltages and currents on each path of an N-path circuit network.
When the network is excited by a time harmonic signal at frequency $\omega_0$, the time-domain relation between the voltage on each path, $v_l\left( t \right)$, is given by \cite{cody}
\begin{equation} \label{eq:td_volt}
{v_l}\left( t \right) = {e^{j{\omega _0}t_0}}{v_{l - 1}}\left( {t - t_0} \right) ,
\end{equation} 
where $t_0=T_s/L$.
The same relation holds for the current on each path. Further, since the system is LPTV, the voltage can be expanded into harmonics of the form \cite{richards}
\begin{equation}
{v_l}\left( t \right) = \sum\limits_{\nu =  - \infty }^\infty  {V_l^\nu {e^{j\left( {{\omega _0} + \nu {\omega _s}} \right)t}}} .
\end{equation} 
Substituting this expansion in (\ref{eq:td_volt}) yields 
\begin{equation} \label{fd_volt}
V_l^\nu = {e^{ - j2\pi \nu / L}}V_{l - 1}^\nu .
\end{equation}
From this expression, it can be seen that the N-path modulation induces a phase shift of $-2 \pi \nu / L $ between the paths for the $\nu^{\rm{th}}$ harmonic.

Next, the interpath relation reported in \cite{cody} for N-path circuit networks will be extended to SD-TWM structures. The implications and applications of the interpath relation will then be examined.

\subsection{Proof of the Interpath Relation}

\begin{figure}
\begin{tikzpicture}
	\definecolor{myred}{rgb}{0.5,0.00,0.5};
	\definecolor{myblue}{rgb}{0.6,1,1};
	\definecolor{mylightblue}{rgb}{.8,1,1};
	\definecolor{mygreen}{rgb}{.6,1,.6};
    \node[anchor=south west,inner sep=0](image)at(0,0,0){\includegraphics[clip,trim=0cm 3cm 0cm 3cm,width=.95\columnwidth]{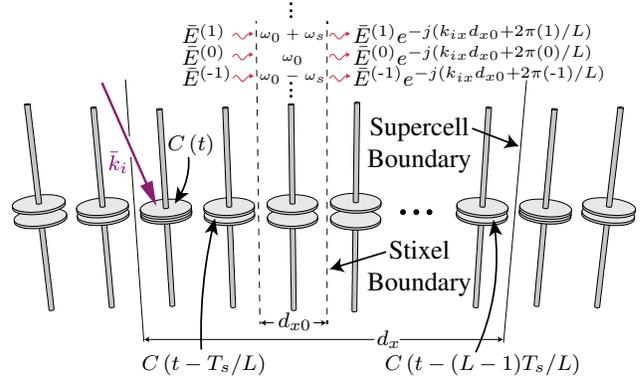}};
    \begin{scope}[x={(image.south east)},y={(image.north west)},font=\fontsize{8}{8}]
		\node[rotate=90] (c) at (.4475,0.805) { ... };
		\node[rotate=90] (c) at (.4475,1.03) { ... };
		\node[black] (c) at  (.4535,.15) {  $d_{x0}$ };
		\node[black] (c) at  (.602,.1) {  $d_{x}$ };
		\node[black] (c) at  (.29,.65) {  $C\left( t \right)$ };
		\node[black] (c) at  (.31,0.03) {  $C\left( t - T_s/L \right)$ };
		\node[black] (c) at  (.75,.03) {  $C\left( t - (L - 1)T_s/L \right)$ };
		\node[align=center,text width=2cm] (c) at (.645,0.46) {{\huge ...}};
		\node[myred] (c) at (.175,0.6) {$\bar k _i$};
		\begin{scope}[font=\fontsize{3}{6}]
			\node[black] (c) at (.45,0.89) {$\omega_0$};
			\node[black] (c) at (.45,0.96) {$\omega_0+\omega_s$};
			\node[black] (c) at (.45,0.84) {$\omega_0-\omega_s$};			
		\end{scope}
		
		\begin{scope}[font=\fontsize{8}{6}]
			\FPset\exa{.31}
			\FPset\exb{.735}
			\FPset\yo{.845}
			\node[black](c)at(\exa,\yo+.06){$\bar E ^{(0)}$};
			\node[black](c)at(\exa,\yo+.12){$\bar E ^{(1)}$};
			\node[black](c)at(\exa+.005,\yo){$\bar E ^{(\mhyphen 1)}$};
			\node[black](c)at(\exb,\yo+.06){$\bar E ^{(0)} e ^ {-j(k_{ix} d_{x0} + 2 \pi (0) / L)}$};
			\node[black](c)at(\exb,\yo+.12){$\bar E ^{(1)} e ^ {-j(k_{ix} d_{x0} + 2 \pi (1) / L)}$};
			\node[black](c)at(\exb+.01,\yo){$\bar E ^{(\mhyphen 1)} e ^ {-j(k_{ix} d_{x0} + 2 \pi (\mhyphen 1) / L)}$};
		\end{scope}
		\node[align=center, text width = 2cm] (c) at (.65,.3) {Stixel Boundary};
		\node[align=center, text width = 2cm] (c) at (.65,.65) {Supercell Boundary};
	\end{scope};
\end{tikzpicture}
\caption{An infinite array of capacitively loaded dipoles modulated in the form of a spatially-discrete traveling-wave. The structure is illuminated by a plane wave at frequency $f_0$ with an incident wavenumber $\bar k _i$. The solid black lines denote the boundaries of a supercell while the dashed lines denote the boundaries of a stixel. The frequency-domain form of the interpath relation derived in Section \ref{sec:relation} is shown relating the frequency harmonics of the fields on the left and right boundaries of a stixel.  }
\label{fig:interpath}
\end{figure}

In this section, the interpath relation \cite{cody_meta} is derived for the case of a plane wave incident upon a SD-TWM structure. As an example, consider the array of capacitively loaded dipoles shown in Fig. \ref{fig:interpath}. The capacitance loading each dipole is a periodic function in time with period $T_s$. Further, the structure is spatially periodic along $x$ with period $d_x$. A single spatial period is referred to as a supercell. Supercells are made up of $L$ sub-domains referred to as stixels, each having a width of $d_{x0}$. The temporal variation of the capacitor within stixel $l$ is related to that within stixel $l-1$ via
\begin{equation} \label{eq:•}
C_l\left( t \right)=C_{l-1}\left(t-t_0\right) ,
\end{equation}
where $t_0=T_s/L$.
From this expression, we observe that the variation of each capacitor is staggered in time, just as in the case of N-path circuit networks. We will now consider two cases of plane wave illumination. In case A, the incident field is given by
\begin{equation} \label{eq:EiA}
\bar {\mathcal{E}}_i^A\left( {\bar r,t} \right) = {\bar E_0}{e^{j\left( {{\omega _0}t - {{\bar k}_i} \cdot \bar r} \right)}} ,
\end{equation}
where ${{\bar k}_i} = {k_{ix}}\hat x + {k_{iy}}\hat y + {k_{iz}}\hat z$.
Meanwhile, in case B, the incident field is given by
\begin{align} \label{eq:EiB}
\begin{split}
\bar {\mathcal{E}}_i^B\left( {x,y,z,t} \right) &= \bar {\mathcal{E}}_i^A\left( {x - {d_{x0}},y,z,t - {t_0}} \right)\\
& = {e^{ - j\left( {{\omega _0}{t_0} - {k_{ix}}{d_{x0}}} \right)}}\bar {\mathcal{E}}_i^A\left( {\bar r,t} \right) .
\end{split} 
\end{align}
We then define the total fields for cases A and B as $\bar {\mathcal{E}} ^ A \left( \bar r, t \right)$ and $\bar {\mathcal{E}} ^B \left( \bar r, t \right)$ respectively. Since the system is linear, we note that
\begin{equation} \label{eq:Eab1}
{{\bar {\mathcal{E}}}^B}\left( {x,y,z,t} \right) = {e^{ - j\left( {{\omega _0}{t_0} - {k_{ix}}{d_{x0}}} \right)}}{{\bar {\mathcal{E}}}^A}\left( {x,y,z,t} \right) .
\end{equation}
Further, by shifting the space-time coordinate system in case B from $(x,y,z,t)$ to $(x',y,z,t')$, such that $x'=x-d_{x0}$ and $t'=t-t_0$, case A is reproduced. This implies
\begin{align} \label{eq:Eab2}
\begin{split}
{{\bar {\mathcal{E}}}^B}\left( {x,y,z,t} \right) &= {{\bar {\mathcal{E}}}^A}\left( {x',y,z,t'} \right)\\
& = {{\bar {\mathcal{E}}}^A}\left( {x - {d_{x0}},y,z,t - {t_0}} \right) .
\end{split}
\end{align}
Equating the right-hand sides of (\ref{eq:Eab1}) and (\ref{eq:Eab2}), we obtain
\begin{equation}
    {e^{ - j\left( {{\omega _0}{t_0} - {k_{ix}}{d_{x0}}} \right)}}{{\bar {\mathcal{E}}}^A}\left( {x,y,z,t} \right) =
    {{\bar {\mathcal{E}}}^A}\left( {x - {d_{x0}},y,z,t - {t_0}} \right) .
\end{equation}
Thus, when a SD-TWM structure is excited by a plane wave, the total fields satisfy
\begin{equation} \label{eq:interpath_td}
\bar {\mathcal{E}}\left( {x,y,z,t} \right) = {e^{j\left( {{\omega _0}{t_0} - {k_{ix}}{d_{x0}}} \right)}}\bar {\mathcal{E}}\left( {x - {d_{x0}},y,z,t - {t_0}} \right) .
\end{equation}
This expression represents the time-domain interpath relation for SD-TWM structures. Since the system is LPTV, the fields throughout space can be expanded into frequency harmonics as \cite{richards}
\begin{equation} \label{eq:•}
\bar {\mathcal{E}}\left( {\bar r,t} \right) = \sum\limits_{\nu  =  - \infty }^\infty  {{{\bar E}^\nu }\left( {\bar r} \right){e^{j\left( {{\omega _0} + \nu {\omega _s}} \right)t}}} .
\end{equation}
Substituting this expansion into (\ref{eq:interpath_td}) yields 
\begin{equation} \label{eq:interpath_fd}
{{\bar E}^\nu }\left( {x,y,z} \right) = {e^{ - j\left( {{k_{ix}}{d_{x0}} + 2\pi \nu /L} \right)}}{{\bar E}^\nu }\left( {x - {d_{x0}},y,z} \right) .
\end{equation}
This expression is the frequency-domain interpath relation for SD-TWM structures.
The interpath relation reveals that, similar to an N-path network, the fields within neighboring stixels at frequency $f_0+\nu f_s$ differ by a phase shift of $-2 \pi \nu / L$ (in addition to the phase shift of the incident wave).
This can be interpreted as an additional frequency-harmonic-dependent tangential momentum imparted by the modulation onto the field. Equation (\ref{eq:interpath_fd}) can also be understood as a modified Floquet boundary condition which accounts for the space-time periodicity of the modulation.

\subsection{Array Interpretation of the Interpath Relation}

\begin{figure}
\begin{tikzpicture}
	\definecolor{myred}{rgb}{0.5,0.00,0.5};
	\definecolor{myblue}{rgb}{0.6,1,1};
	\definecolor{mylightblue}{rgb}{.8,1,1};
	\definecolor{mygreen}{rgb}{.6,1,.6};
    \node[anchor=south west,inner sep=0](image)at(0,0,0){\includegraphics[clip,trim=0cm 2cm 0cm 5cm,width=.95\columnwidth]{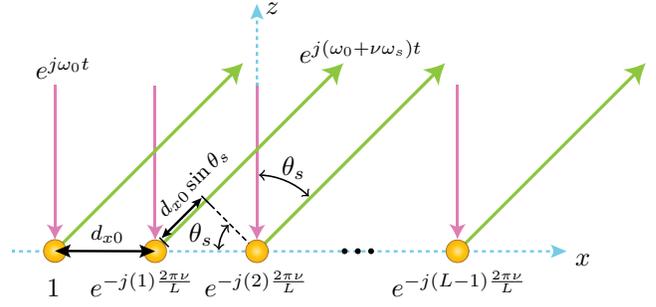}};
    \begin{scope}[x={(image.south east)},y={(image.north west)},font=\fontsize{10}{8}]
        \FPset\xo{.065}
        \FPset\yo{.1}
        \node[align=center] (c) at (\xo,\yo-.015) {$1$};
        \node[align=center] (c) at (\xo+.14,\yo) {$e^{-j(1)\frac{2\pi \nu}{L}}$};
        \node[align=center] (c) at (\xo+2*.16,\yo) {$e^{-j(2)\frac{2\pi \nu}{L}}$};
        \node[align=center] (c) at (\xo+4*.16,\yo) {$e^{-j(L-1)\frac{2\pi \nu}{L}}$};
        \node[align=center] (c) at (\xo+3*.16,.197) {{\huge ...}};
        \node[align=center] (c) at (.08,.75) {$e^{j \omega_0 t}$};
        \node[align=center] (c) at (.55,.8) {$e^{j \left(\omega_0 + \nu \omega_s \right)t}$};
        \node[align=center] (c) at (.9,.17) {$x$};        
        \node[align=center] (c) at (.41,.93) {$z$};        
        \begin{scope}[shift={(.385,.2)}]
	        \coordinate (o) at (0,0);
	        \coordinate (z) at (89:.1);
	        \coordinate (a) at (63:.1);
	        \coordinate (x) at (179:.1);
	        \coordinate (b) at (120:.1);
	        \draw pic [draw=black,fill=none,angle radius=10mm,semithick,<->] {angle = a--o--z} node[above=31pt,right=6pt] {$\theta_s$};
	        \draw pic [draw=black,fill=none,angle radius=5mm,semithick,<->] {angle = b--o--x} node[above=6pt,left=13pt] {$\theta_s$};	        
        \end{scope};     
        \begin{scope}[font=\fontsize{8}{6}]
	        \node[align=center] (c) at (.15,.25) {$d_{x0}$};  
	        \node[align=center,rotate=45] (c) at (.285,.405) {$d_{x0}\sin \theta_s$};  
        \end{scope};        
	\end{scope};
\end{tikzpicture}
\caption{Array interpretation of the scattering due to a single supercell consisting of $L$ stixels for a given observation angle, $\theta_s$, and observation frequency $f_0 + \nu f_s$. In this example, the monochromatic illumination at frequency $f_0$ is assumed to be normally incident ($k_{ix}=k_{iy}=0$). Therefore, the relative phase of each source is given soley by the phase imparted by the modulation.} \label{fig:array}
\end{figure}

The interpath relation in (\ref{eq:interpath_fd}) provides insight into the scattering behavior of SD-TWM structures. Let us consider the array factor produced by a single supercell of scatterers, as shown in Fig. \ref{fig:array} \cite{eucap}. From (\ref{eq:interpath_fd}), the array factor for frequency $f_0 + \nu f_s$ at observation positions in the $x$-$z$ plane is given by
\begin{equation} \label{eq:af}
{\rm{ARFAC}} \left(\theta_s\right) = \sum\limits_{l = 0}^{L - 1} {{e^{jl\left( { - {k_{ix}}d_{x0} - \frac{{2\pi \nu }}{L} + \frac{\omega_0+\nu \omega_s}{c} d_{x0}\sin {\theta _s}} \right)}}} ,
\end{equation}
where $\theta_s$ is the observation angle with respect to the $z$ axis. The relative phase of the excitation on each scatterer is determined by the incident angle and the particular frequency of observation. The inter-element phase due to the incident wave is $k_{ix}d_{x0}$. Meanwhile, the inter-element phase induced by the SD-TWM is $2 \pi \nu / L$, where $\nu$ corresponds to the observed frequency of $f_0 + \nu f_s$.

For simplicity, let us focus on the array factor when the excitation is normally incident ($k_{ix}=k_{iy}=0$). In this case, the array factor can be written as
\begin{equation} \label{eq:af_normal}
{\rm{ARFAC}} \left(\theta_s\right) = \sum\limits_{l = 0}^{L - 1} {{e^{jl\left( { - \frac{{2\pi \nu }}{L} + \frac{\omega_0+\nu \omega_s}{c} d_{x0}\sin {\theta _s}} \right)}}}  .
\end{equation}
From antenna array theory and (\ref{eq:af_normal}), the beam-pointing direction(s) is dependent on the frequency harmonic and satisfies \cite{balanis}
\begin{equation} \label{eq:beam_pointing}
\frac{\omega_0+\nu \omega_s}{c} d_{x0}\sin {\theta _s} = \frac{2\pi\nu}{L} + 2\pi p ,
\end{equation}
where $p$ is an integer. If the modulation frequency is much smaller than the RF carrier frequency, then $\omega_0+\nu\omega_s \approx \omega_0$. In this case, when $\nu$ is replaced by $\nu + L$, the beam-pointing angle(s) remains the same (since we are free to select $p$). This reveals that frequencies separated by $p' L f_s$ ($p' \in \mathbb{Z}$) are scattered to the same angle (or multiple angles). This effect, shown pictorially in Fig. \ref{fig:fancy}, is not predicted when idealized continuous traveling-wave modulation is assumed \cite{oli}. Equation (\ref{eq:beam_pointing}) also reveals that if  
\begin{equation} \label{eq:•}
\frac{\omega_0+\nu \omega_s}{c}d_{x0} < \left|\frac{2\pi\nu}{L}+2\pi p\right|, \quad \forall p \in \mathbb{Z} ,
\end{equation}
then frequency $f_0+\nu f_s$ does not correspond to any propagating angles. Frequencies which satisfy this condition are bound to the nearfield of the metasurface. A further discussion on the scattering behavior of SD-TWM is provided in \cite{zca}.

\subsection{Applications of the Interpath Relation}

The interpath relation provides valuable insight into the behavior of SD-TWM structures. It reveals the potential for these structures to achieve unique electromagnetic phenomena, such as sub-harmonic mixing and retro-reflective frequency conversion, as explored in \cite{zca}. By appropriately tailoring the time-dependence of a SD-TWM structure (e.g. Fig. \ref{fig:fancy}), the frequency of the reflected wave can be shifted by an integer multiple of $L f_s$, where $L$ is the number of stixels per period. This effect is known as sub-harmonic mixing since the reflected wave is shifted by an integer multiple of $f_s$. Further, antenna array theory and (\ref{eq:af}) can be used to compute the required angle of incidence and $L$ such that the scattered wave reflects back toward the illumination source at a translated frequency (i.e. retro-reflective frequency conversion). Finally, for structures designed theoretically via a continuous traveling-wave model, (\ref{eq:beam_pointing}) can be used to find the maximum unit cell size such that grating lobes are not produced at any frequency in the physical  implementation. This is particularly important when the modulation frequency, $f_s$, is comparable to the RF carrier frequency, $f_0$.

In addition to providing physical understanding, the interpath relation can be used in practice to enable the simulation of SD-TWM structures. For both physical and idealized models, this relation will be shown in Section \ref{sec:mom} to dramatically reduce the number of unknowns required to simulate these structures. The reduction in unknowns comes at no cost to the accuracy of the simulation since the interpath relation contains no approximations. For complicated unit cell designs which require a high level of discretization, the ability to simulate a single stixel can be a dramatic improvement over the simulation of several time-varying unit cells simultaneously. Further, idealized designs obtained via a continuous traveling-wave modulation model (such as those discussed in \cite{hyper,fdtd,nonrecip,stdiff_grat,optical_leakywave,st_metasurf_alu}) can also benefit from a reduction in the number of unknowns required for simulation. This is because the stixel size for these structures can be made arbitrarily small. Without using the interpath relation, numerically simulating a continuous traveling-wave modulated structure would require the space-time variation to be discretized into individual time-varying computational elements over an entire spatial period \cite{fdtd2,caloz}. Using the interpath relation, we can shrink the computational domain to the width of a single computational element. This reduces the number of unknowns by the number of elements used in the simulation without the interpath relation.

\section{Method of Moments Formulation using the Interpath Relation} \label{sec:mom}

In this section we will derive the MoM formulation for a representative SD-TWM structure. We will begin by deriving the MoM formulation for the time-invariant structure shown in Fig. \ref{fig:lti_struc}. The formulation for the time-invariant case will then be modified to account for SD-TWM. Specifically, in Section \ref{sec:mom_LTI}, the MoM formulation will be derived for a periodic, inhomogeneous, time-invariant impedance sheet placed above a grounded dielectric substrate. In Section \ref{sec:mom_trav_wave}, the time-invariant MoM formulation from Section \ref{sec:mom_LTI} will be modified to account for a SD-TWM impedance sheet.

\subsection{Inhomogeneous Time-Invariant Structure} \label{sec:mom_LTI}

\begin{figure}[!t]
\centering
\FPset\lenax{0.45}
\FPset\lenay{0.3}
\FPset\lenaz{0.4}
\FPset\lenk{0.35}
\FPset\angkd{75}
\FPmul\angkr\angkd{0.0175}
\FPset\angpd{10}
\FPmul\angpr\angpd{0.0175}
\FPcos\ksx\angkr
\FPsin\ksy\angkr
\FPtan\ptan\angpr
\FPset\lent{.01}
\FPset\rx{-35}
\FPset\ry{38}
\FPset\hh{.05}
\FPset\gap{.065}
\FPset\xh{.375}
\FPset\yh{.24}
\begin{tikzpicture}
	\definecolor{myred}{rgb}{0.40,0.00,0.60};
	\definecolor{myblue}{rgb}{0,.4,1};
	\definecolor{mygreen}{rgb}{0,.7,.4};
    \node[anchor=south west,inner sep=0](image)at(0,0,0){\includegraphics[clip,trim=0cm 3cm 0cm 3cm,width=.95\columnwidth]{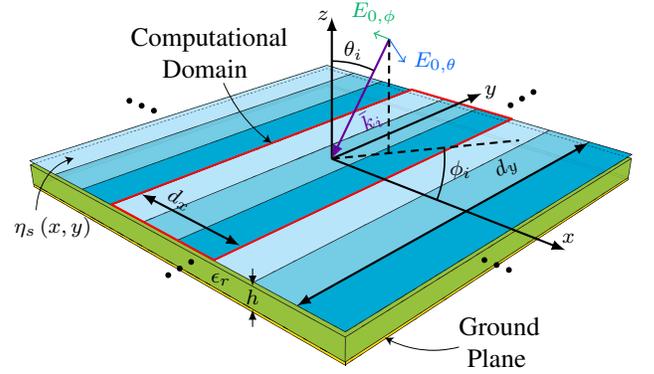}};
    \begin{scope}[x={(image.south east)},y={(image.north west)},font=\fontsize{8}{6}]
        \begin{scope}[shift={(0.5,0.6)}] 
          \draw[black,thick,-latex] coordinate (o) (0,0)  -- coordinate (x) (\rx:\lenax) node[above=4.5pt,right=-5pt] {$x$};
          \draw[black,thick,-latex] (o) -- coordinate[pos=1] (y)(\ry:\lenay) node[above=.5pt,right=-2.5pt] {$y$};
          \draw[black,thick,-latex] (o) -- coordinate[pos=1] (z)(90:\lenaz) node[above=.5pt,left=-2.5pt] {$z$};
          \draw[myred,thick,latex-] (o) -- coordinate[pos=1] (k)(\angkd:\lenk) node[midway,rotate=30,below=9pt,right=-8pt] {$\bar k_i$};
          \draw[black,thick,densely dashed] (k) -- coordinate[pos=1] (p) (\ksx*\lenk,\ksx*\lenk*\ptan);
          \draw[black,thick,densely dashed] (o) -- (\angpd:.3);
          \draw[myblue,->] [shift={(\angkd:\lenk)}] (0,0) -- (-70:.07) node[right] {$E_{0,\theta}$};
          \draw[mygreen,->] [shift={(\angkd:\lenk)}] (0,0) -- (180-35:.03) node[above] {$E_{0,\phi}$};
          \draw pic [draw=black,fill=none,angle radius=13mm,semithick] {angle = k--o--z} node[above=41pt,right=1pt] {$\theta_i$};
          \draw pic [draw=black,fill=none,angle radius=15mm,semithick] {angle = x--o--p} node[below=4pt,right=41pt] {$\phi_i$};
        \end{scope}
        
        \draw[black,thick,latex-latex] (.2,.5)  -- (.355,.355) node[near start,sloped,above=5pt,right=-5pt] {$d_x$};
        \draw[black,thick,latex-latex] (.445,.185)  -- (.905,.645) node[near end, sloped,above=-.5pt] {$d_{y}$};
        
        \begin{scope}[shift={(\xh,\yh)}]
	        \draw[black,-latex] (0,\hh)  -- (0,0);
	        \draw[black,-latex] (0,-\hh-\gap)  -- (0,-\gap);
	        \node[align=center] (c) at (0,-0.45*\gap) {$h$};
	        \node[align=center] (c) at (-.05,.02) {$\epsilon_r$};
        \end{scope}
		
		\node[align=center,text width=2cm, text=black] (c) at (0.06,.405) {$\mathclap{\eta_s \left(x,y\right)}$};
		\node[align=center,text width=1cm, text=black] (c) at (0.76,.1) {{\fontsize{12}{2} Ground Plane}};
		\node[align=center,text width=3cm, text=black] (c) at (0.3,.9) {{\fontsize{12}{2} Computational Domain}};		
		
		\node[align=center,text width=2cm, text=black,rotate=-22] (c) at (.76,0.3) {{\huge ...}}; 
		\node[align=center,text width=2cm, text=black,rotate=-22] (c) at (.2,0.75) {{\huge ...}}; 
		\node[align=center,text width=2cm, text=black,rotate=30] (c) at (.26,0.29) {{\huge ...}}; 
		\node[align=center,text width=2cm, text=black,rotate=30] (c) at (.8,0.77) {{\huge ...}}; 
    \end{scope}
\end{tikzpicture}
\caption{An illustration of a plane wave incident upon an inhomogeneous impedance sheet placed over a grounded dielectric. Since the structure is periodic in space, the computational domain contains a single spatial period. It should be noted that this is just one example of a 2D spatially periodic impedance sheet. The presented formulation is valid for periodic variations along $y$, although this is not depicted in the figure.}
\label{fig:lti_struc}
\end{figure}

Here, we will derive the MoM formulation for the time-invariant structure shown in Fig. \ref{fig:lti_struc} under a plane-wave illumination.
The isotropic sheet impedance placed on the surface of the grounded dielectric, ${\eta_s}\left( {x,y} \right)$, is periodic in $x$ with period $d_x$ and in $y$ with period $d_y$. The dielectric has a relative permittivity of ${\epsilon _r}$ and a thickness of $h$. The excitation field phasor is a plane wave of the form
\begin{align} \label{eq:inc}
\begin{split}
{\bar E_i}\left( {\bar r} \right) &= {\bar E_0}{e^{ - j{{\bar k}_i} \cdot \bar r}} \\
&={\bar E_0}{e^{ - j\left( {{k_{ix}}x + {k_{iy}}y + {k_{iz}}z} \right)}} \\
 &= {\bar E_0}{e^{j{k_0}\left( {\sin {\theta _i}\cos {\phi _i}x + \sin {\theta _i}\sin {\phi _i}y + \cos {\theta _i}z} \right)}} .
\end{split}
\end{align}
The procedure used here to obtain the MoM matrix equation closely follows the formulation presented by Jin in \cite{jin} for simulating a planar array of metallic patches.


First, we will write the boundary condition for the total field on the impedance sheet and subsequently express the scattered electric field in terms of the surface current density. The boundary condition at $z=0$ is 
\begin{equation} \label{eq:BC1}
{\bar E_t }\left( {x,y} \right) = {\eta_s}\left( {x,y} \right){\bar J_s}\left( {x,y} \right) ,
\end{equation}
where ${\bar E_t }\left( {x,y} \right)$ denotes the transverse component of the total electric field in the $x$-$y$ plane. The total electric field can be separated into the incident field produced in the absence of the impedance sheet and the field scattered by the currents induced on the impedance sheet. Note that the incident field in the absence of the impedance sheet contains both the excitation field and the field reflected from the grounded dielectric. On the surface of the grounded dielectric, the reflected electric field can be written as 
\begin{equation} \label{eq:•}
{{\bar E}_{r,t}}\left( {x,y} \right) = \mathord{\buildrel{\lower3pt\hbox{$\scriptscriptstyle\leftrightarrow$}} 
\over \Gamma } {{\bar E}_{i,t}}\left( {x,y} \right) ,
\end{equation}
where $\mathord{\buildrel{\lower3pt\hbox{$\scriptscriptstyle\leftrightarrow$}} 
\over \Gamma }$ is the dyadic reflection coefficient which is a function of $\theta_i$, $\phi_i$, $h$ and $\epsilon_r$. By splitting the total electric field into incident and scattered components, (\ref{eq:BC1}) can be rewritten as
\begin{equation} \label{eq:BC_LTI}
\left[ {1 + \mathord{\buildrel{\lower3pt\hbox{$\scriptscriptstyle\leftrightarrow$}} 
\over \Gamma } } \right] {\bar E_{i,t}}\left( {x,y} \right)
+{\bar E_{s,t}}\left( {x,y} \right)
={\eta_s}\left( {x,y} \right){\bar J_s}\left( {x,y} \right) ,
\end{equation}
where ${\bar E_{i,t}}$ and ${\bar E_{s,t}}$ denote the transverse components of the excitation and scattered electric field respectively. Since the structure is periodic in $x$ and $y$, it follows from Floquet's Theorem that
\begin{equation} \label{eq:Floq_Thrm}
{\bar J_s}\left( {x,y} \right) = {e^{ - j\left( {{k_{ix}}x + {k_{iy}}y} \right)}}{\bar j_s}\left( {x,y} \right) ,
\end{equation}
such that ${\bar j_s}\left( {x,y} \right)$ is periodic in $x$ with period $d_x$ and in $y$ with period $d_y$. Since ${\bar j_s}\left( {x,y} \right)$ is periodic in $x$ and $y$, it can be expanded in terms of a 2D Fourier series as
\begin{equation} \label{eq:•}
{\bar j_s}\left( {x,y} \right) = \sum\limits_{p =  - \infty }^\infty  {\sum\limits_{q =  - \infty }^\infty  {{{\bar I}_{pq}}{e^{-j\left( {{k'_{xp}}x + {k'_{yq}}y} \right)}}} } ,
\end{equation}
where ${k'_{xp}} = {\textstyle{{2\pi p} \over d_x}}$ and ${k'_{yq}} = {\textstyle{{2\pi q} \over d_y}}$. Substituting this into (\ref{eq:Floq_Thrm}), we obtain 
\begin{equation} \label{eq:curr_exp}
{\bar J_s}\left( {x,y} \right) = \sum\limits_{p =  - \infty }^\infty  {\sum\limits_{q =  - \infty }^\infty  {{{\bar I}_{pq}}{e^{-j\left( {{k_{xp}}x + {k_{yq}}y} \right)}}} } ,
\end{equation}
where ${k_{xp}} = {k_{ix}} + {k'_{xp}}$ and ${k_{yq}} = {k_{iy}} + {k'_{yq}}$. We can interpret (\ref{eq:curr_exp}) as an expansion of the surface current density into sheet currents with uniform amplitude $\bar I _{pq}$ and a phase dependence similar to a plane wave.
Therefore, at $z=0$, the transverse scattered electric field can be written as 
\begin{multline} \label{eq:escat}
{\bar E_{s,t}}\left( {x,y} \right) = \\ 
 - j{k_0}{Z_0}\sum\limits_{p  - \infty }^\infty  {\sum\limits_{q =  - \infty }^\infty  {\mathord{\buildrel{\lower3pt\hbox{$\scriptscriptstyle\leftrightarrow$}} 
\over G} \left( {{k_{xp}},{k_{yq}}} \right){{\bar I}_{pq}}{e^{ - j\left( {{k_{xp}}x + {k_{yq}}y} \right)}}} }   ,
\end{multline}
where ${\mathord{\buildrel{\lower3pt\hbox{$\scriptscriptstyle\leftrightarrow$}} 
\over G} }\left( {{k_{xp}},{k_{yq}}} \right)$ is the spectral-domain representation of the dyadic Green's function as derived in Appendix \ref{sec:A1}. Substituting the expansion for the scattered electric field into (\ref{eq:BC_LTI}) and multiplying both sides by ${e^{ j\left( {{k_{ix}}x + {k_{iy}}y} \right)}}$, we obtain
\begin{align} \label{eq:integral_eqn}
\begin{split}
\left[ {1 + \mathord{\buildrel{\lower3pt\hbox{$\scriptscriptstyle\leftrightarrow$}} 
\over \Gamma } } \right]{\bar E_{0,t}} &= {\eta_s}\left( {x,y} \right){\bar j_s}\left( {x,y} \right) \\
&+j k_0 Z_0 \sum\limits_{p =  - \infty }^\infty  {\sum\limits_{q =  - \infty }^\infty  {{{\mathord{\buildrel{\lower3pt\hbox{$\scriptscriptstyle\leftrightarrow$}} 
\over G} }_{pq}}{{\bar I}_{pq}}{e^{ - j\left( {{k'_{xp}}x + {k'_{yq}}y} \right)}}} } ,
\end{split}
\end{align}
where $\bar E_{0,t}$ is the transverse component of the excitation field amplitude, $\bar E_0$, and ${{{\mathord{\buildrel{\lower3pt\hbox{$\scriptscriptstyle\leftrightarrow$}} 
\over G} }_{pq}}} = {\mathord{\buildrel{\lower3pt\hbox{$\scriptscriptstyle\leftrightarrow$}} 
\over G} }\left( {{k_{xp}},{k_{yq}}} \right)$. As discussed in \cite{jin}, (\ref{eq:integral_eqn}) represents our integral equation since $\bar I_{pq}$ is given by
\begin{equation} \label{eq:2dft}
{{\bar I}_{pq}} = \frac{1}{{d_x d_y}}\int\limits_{ - d_x/2}^{d_x/2} {\int\limits_{ - d_y/2}^{d_y/2} {{{\bar j}_s}\left( {x,y} \right){e^{j\left( {{k'_{xp}}x + {k'_{yq}}y} \right)}}dxdy} } .
\end{equation}

\begin{figure}[!t]
\centering
\FPset\lenax{0.2}
\FPset\lenay{0.185}
\FPset\lenaz{0.25}
\FPset\lent{.01}
\FPset\rx{-35}
\FPset\ry{40}
\begin{tikzpicture}
    \node[anchor=south west,inner sep=0](image)at(0,0,0){\includegraphics[clip,trim=0cm 3cm 0cm 3cm,width=.95\columnwidth]{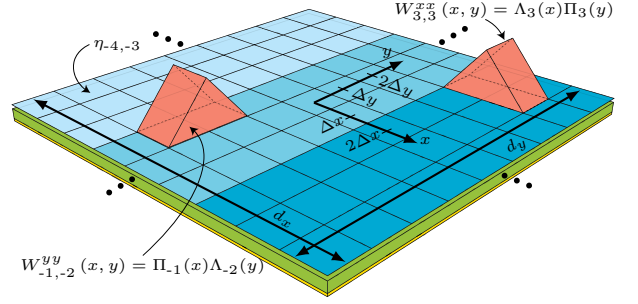}};
    \begin{scope}[x={(image.south east)},y={(image.north west)},font=\fontsize{6}{6}]
        \begin{scope}[shift={(0.5,0.6)}] 
          \draw[black,thin,shift={(\rx:0.07)}] (180+\ry:\lent) -- (\ry:\lent) node[rotate=25,below=-1pt,left=1pt] {$\Delta x$};
          \draw[black,thin,shift={(\rx:2*0.07)}] (180+\ry:\lent) -- (\ry:\lent) node[rotate=25,below=-1pt,left=1pt] {$2\Delta x$};
          \draw[black,thin,shift={(\ry:0.06)}] (180+\rx:\lent) -- (\rx:\lent) node[rotate=-20,above=0pt,right=-2pt] {$\Delta y$};
          \draw[black,thin,shift={(\ry:2*0.06)}] (180+\rx:\lent) -- (\rx:\lent) node[rotate=-20,above=-.9pt,right=-2.7pt] {$2\Delta y$};
          \draw[black,thick,-latex] coordinate (o) (0,0)  -- coordinate (a) (\rx:\lenax) node[below=-1pt,right=-3pt] {$x$};
          \draw[black,thick,-latex] (o) -- coordinate (c)(\ry:\lenay) node[above=3pt,right=-12pt] {$y$};
        \end{scope}
        
        \draw[black,thick,latex-latex] (.057,.61)  -- (.55,.15) node[near end,sloped,above=4pt,right=-3pt] {$d_x$};
        \draw[black,thick,latex-latex] (.47,.165)  -- (.92,.635) node[near end, sloped,below=-1pt] {$d_{y}$};
		
		\node (c) at (0.23,.14) {$W_{\mhyphen 1,\mhyphen 2}\cup{yy}\left(x,y\right)=\Pi_{\mhyphen 1}(x)\Lambda_{\mhyphen 2}(y)$};
		\node (c) at (0.8,.86) {$W_{3,3}\cup{xx}\left(x,y\right)=\Lambda_{3}(x)\Pi_{3}(y)$};
		\node (c) at (0.19,.76) {$\mathclap{\eta_{\mhyphen 4,\mhyphen 3}}$};
		
		\node[align=center,text width=2cm, text=black,rotate=-22] (c) at (.82,0.38) {{\huge ...}};
		\node[align=center,text width=2cm, text=black,rotate=-22] (c) at (.27,0.78) {{\huge ...}};
		\node[align=center,text width=2cm, text=black,rotate=30] (c) at (.195,0.365) {{\huge ...}};
		\node[align=center,text width=2cm, text=black,rotate=30] (c) at (.72,0.79) {{\huge ...}};
    \end{scope}
\end{tikzpicture}
\caption{An overlay of the basis functions used to expand the current within a spatial period of the structure shown in Fig. \ref{fig:lti_struc}. In this example, there are $2M'+1=9$ unknowns along $x$ and $2N'+1=9$ unknowns along $y$}
\label{fig:basis}
\end{figure}

Numerically computing the surface current requires us to expand ${{\bar j}_s}\left( {x,y} \right)$ into a set of basis functions. Following the procedure in \cite{jin}, we will expand ${{\bar j}_s}\left( {x,y} \right)$ into basis functions with finite divergence. The basis function profile for the $x$ component of  ${\bar j}_s$, $W_{mn}^{xx}\left( {x,y} \right)$, and the $y$ component of  ${\bar j}_s$, $W_{mn}^{yy}\left( {x,y} \right)$, are shown in Fig. \ref{fig:basis}. $W_{mn}^{xx}\left( {x,y} \right)$ is the product of a rooftop function, ${\Lambda _m}$, in $x$ and a pulse function, ${\Pi _n}$, in $y$. Meanwhile, $W_{mn}^{yy}\left( {x,y} \right)$ is the product of a rooftop function, ${\Lambda _n}$, in $y$ and a pulse function, ${\Pi _m}$, in $x$. 
\begin{equation} \label{eq:Wxx}
W_{mn}^{xx}\left( {x,y} \right) = {\Lambda _m}\left( x \right){\Pi _n}\left( y \right)
\end{equation}
\begin{equation} \label{eq:Wyy}
W_{mn}^{yy}\left( {x,y} \right) = {\Pi _m}\left( x \right){\Lambda _n}\left( y \right)
\end{equation}
The explicit form of these basis function is provided here as a reference. Assuming there are $M=2M'+1$ spatial unknowns along $x$ and $N=2N'+1$ spatial unknowns along $y$,
\begin{equation} \label{eq:pix}
{\Pi _m}\left( x \right) = \left\{ {\begin{array}{*{20}{l}}
{1,\quad x \in \left[ {\left( {m - {\textstyle{1 \over 2}}} \right)\Delta x,\left( {m + {\textstyle{1 \over 2}}} \right)\Delta x} \right]}\\
{0,\quad {\rm{otherwise}}}
\end{array}} \right. 
\end{equation}
\begin{equation} \label{eq:piy}
{\Pi _n}\left( y \right) = \left\{ {\begin{array}{*{20}{l}}
{1,\quad y \in \left[ {\left( {n - {\textstyle{1 \over 2}}} \right)\Delta y,\left( {n + {\textstyle{1 \over 2}}} \right)\Delta y} \right]}\\
{0,\quad {\rm{otherwise}}}
\end{array}} \right. 
\end{equation} 
\begin{multline} \label{eq:lamx}
{\Lambda _m}\left( x \right) = \\
\left\{ {\rar{1.2}  \begin{array}{*{20}{l}}
{{\textstyle{3 \over 2}} + {\textstyle{x \over {\Delta x}}}-m, \texttt{ } x \in \left[ {\left( {m - {\textstyle{3 \over 2}}} \right)\Delta x,\left( {m - {\textstyle{1 \over 2}}} \right)\Delta x} \right]}\\
{{\textstyle{1 \over 2}} - {\textstyle{x \over {\Delta x}}}+m, \texttt{ } x \in \left[ {\left( {m - {\textstyle{1 \over 2}}} \right)\Delta x,\left( {m + {\textstyle{1 \over 2}}} \right)\Delta x} \right]}\\
{0,\quad {\rm{otherwise}}}
\end{array}} \right. 
\end{multline}
\begin{multline} \label{eq:lamy}
{\Lambda _n}\left( y \right) = \\
 \left\{ {\rar{1.2} \begin{array}{*{20}{l}}
{{\textstyle{3 \over 2}} + {\textstyle{y \over {\Delta y}}}-n, \texttt{ } y \in \left[ {\left( {n - {\textstyle{3 \over 2}}} \right)\Delta y,\left( {n - {\textstyle{1 \over 2}}} \right)\Delta y} \right]}\\
{{\textstyle{1 \over 2}} - {\textstyle{y \over {\Delta y}}}+n, \texttt{ } y \in \left[ {\left( {n - {\textstyle{1 \over 2}}} \right)\Delta y,\left( {n + {\textstyle{1 \over 2}}} \right)\Delta y} \right]}\\
{0,\quad {\rm{otherwise}}}
\end{array}} \right. ,
\end{multline} 
where $\Delta x = d_x/M$ and $\Delta y = d_y/N$.  The surface current can therefore be compactly represented as
\begin{equation} \label{eq:basis} \renewcommand{\arraystretch}{1}
{{\bar j}_s}\left( {x,y} \right) = \sum\limits_{m' =  - M'}^{M'} {\sum\limits_{n' = -N'}^{N'} {{{\mathord{\buildrel{\lower3pt\hbox{$\scriptscriptstyle\leftrightarrow$}} 
\over W} }_{m'n'}}\left( {x,y} \right){{\bar j}_{m'n'}}} } ,
\end{equation}
where
\begin{equation} \label{eq:W}
\tensor{W}_{mn}\left(x,y\right)=\Lambda_m\left(x\right)\Pi_n\left(y\right) \hat x \hat x +\Pi_m\left(x\right)\Lambda_n\left(y\right) \hat y \hat y .
\end{equation}
We can substitute (\ref{eq:basis}) into (\ref{eq:2dft}) to obtain $\bar I_{pq}$ in terms of the weighting coefficients ${{\bar j}_{m'n'}}$, which yields
\begin{equation} \label{eq:Ipq}
{{\bar I}_{pq}} = \frac{1}{{MN}}{\tensor{T}_{pq}}\sum\limits_{m'n'} {{{\bar j}_{m'n'}}{e^{j\left( {2\pi pm'/M + 2\pi qn'/N} \right)}}} ,
\end{equation}
where ${{\mathord{\buildrel{\lower3pt\hbox{$\scriptscriptstyle\leftrightarrow$}} 
\over T} }_{pq}}$ represents the 2D Fourier series coefficients of ${{\mathord{\buildrel{\lower3pt\hbox{$\scriptscriptstyle\leftrightarrow$}} 
\over W} }_{00}}$ given by
\begin{align} \label{eq:Tpq}
\begin{split}
\tensor{T}_{pq}  = & {\rm sinc} ^2 \left( \frac{p\pi}{M} \right) {\rm sinc} \left( \frac{q\pi}{N} \right)e^{j\frac{p\pi}{M}} \hat x \hat x 
\\ + & {\rm sinc} \left( \frac{p\pi}{M} \right) {\rm sinc} ^2 \left( \frac{q\pi}{N} \right)e^{j\frac{q\pi}{N}} \hat y \hat y 
\end{split} .
\end{align}
We now substitute ${\bar j}_s$ from (\ref{eq:basis}) and ${\bar I}_{pq}$ from (\ref{eq:Ipq}) into (\ref{eq:integral_eqn}) to obtain an expression relating the incident field amplitude to the unknown current weighting coefficients.
\begin{align} \label{eq:pretest}
\begin{split}
&\left[ {1 + \mathord{\buildrel{\lower3pt\hbox{$\scriptscriptstyle\leftrightarrow$}} 
\over \Gamma } } \right]{{\bar E}_{0,t}} = \sum\limits_{m'n'} {{\eta_s}\left( {x,y} \right)\mathord{\buildrel{\lower3pt\hbox{$\scriptscriptstyle\leftrightarrow$}} 
\over W} _{m'n'} \left( {x,y} \right){{\bar j}_{m'n'}}} \\
 & + j\frac{{{k_0}{Z_0}}}{{MN}} \sum\limits_{m'n'} {\sum\limits_{pq} {{e^{ - j\left( {{k'_{xp}}x + {k'_{yq}}y} \right)}}{{\mathord{\buildrel{\lower3pt\hbox{$\scriptscriptstyle\leftrightarrow$}} 
\over G} }_{pq}}{{\mathord{\buildrel{\lower3pt\hbox{$\scriptscriptstyle\leftrightarrow$}} 
\over T} }_{pq}} H_{m'n'}^{pq} {{\bar j}_{m'n'}}} } 
\end{split} ,
\end{align}
where 
$H_{m'n'}^{pq}={{e^{j\left( {2\pi pm'/M + 2\pi qn'/N} \right)}}}$ is the phase term from (\ref{eq:Ipq}).

The final MoM matrix equation can be obtained by testing the left- and right-hand sides of (\ref{eq:pretest}) with 
\begin{multline} \label{eq:•}
\frac{1}{{d_x d_y}}\int\limits_{ - d_x/2}^{d_x/2} {\int\limits_{ - d_y/2}^{d_y/2} {{\tensor{W}_{mn}}\left( {x,y} \right)\left\{  \bullet  \right\}dxdy} } \\
\forall \left\{ {m \in \left[ { - M',M'} \right],n \in \left[ { - N',N'} \right]} \right\} .
\end{multline}
Here, we have chosen to use the Galerkin method, i.e., the testing functions are the same as the basis functions. The testing operation is carried out for all observation positions $(m \Delta x,n \Delta y )$ within the spatial period. To obtain an explicit form for the entries of the MoM matrix, we will approximate the surface impedance as a summation over pulse functions.
\begin{equation} \label{eq:Zs}
{\eta_s}\left( {x,y} \right) = \sum\limits_{m'' =  - M'}^{M'} {\sum\limits_{n'' =  - N'}^{N'} {{\Pi _{m''}}\left( x \right){\Pi _{n''}}\left( y \right)\eta_{m''n''}} } 
\end{equation}
Carrying out the integrations, we obtain the final MoM matrix equation, which can be written as
\begin{align} \label{eq:LTIMOM}
\begin{split}
\left[ {1 + \mathord{\buildrel{\lower3pt\hbox{$\scriptscriptstyle\leftrightarrow$}} 
\over \Gamma } } \right]{{\bar E}_{0,t}} & = \sum\limits_{m'n'} {\left( {\mathord{\buildrel{\lower3pt\hbox{$\scriptscriptstyle\leftrightarrow$}} 
\over \eta_{mn,m'n'}}  + {{\mathord{\buildrel{\lower3pt\hbox{$\scriptscriptstyle\leftrightarrow$}} 
\over Z} }_{m - m',n - n'}}} \right){{\bar j}_{m'n'}}} \\
& \quad \forall \left\{ {m \in \left[ { - M',M'} \right],n \in \left[ { - N',N'} \right]} \right\}
\end{split} ,
\end{align}
where
\begin{align} \label{eq:Zxx}
\begin{split}
\hat x \cdot & \tensor{\eta} _{mn,m'n'} \cdot \hat x  = \\
& \quad {\delta _{n-n'}}\left\{ {\rar{1.2} \begin{array}{*{20}{l}}
{{\textstyle{1 \over 6}}\eta_{m - 1,n},\quad m' = m - 1}\\
{{\textstyle{1 \over 3}}\left\{ {\eta_{m - 1,n} + \eta_{m,n}} \right\},\quad m' = m}\\
{{\textstyle{1 \over 6}}\eta_{m,n},\quad m' = m + 1}\\
{0,\quad {\rm{otherwise}}}
\end{array}} \right.
\end{split}
\end{align}
\begin{align} \label{eq:Zyy}
\begin{split}
\hat y \cdot & \tensor{\eta} _{mn,m'n'} \cdot \hat y = \\
& \quad {\delta _{m-m'}}\left\{ {\rar{1.2} \begin{array}{*{20}{l}}
{{\textstyle{1 \over 6}}\eta_{m,n - 1},\quad n' = n - 1}\\
{{\textstyle{1 \over 3}}\left\{ {\eta_{m,n - 1} + \eta_{m,n}} \right\},\quad n' = n}\\
{{\textstyle{1 \over 6}}\eta_{m,n},\quad n' = n + 1}\\
{0,\quad {\rm{otherwise}}}
\end{array}} \right.
\end{split}
\end{align}
\begin{equation} \label{eq:Zxy}
\hat x \cdot \tensor{\eta}_{mn,m'n'} \cdot \hat y = \hat y \cdot \tensor{\eta}_{mn,m'n'} \cdot \hat x = 0
\end{equation}
\begin{equation} \label{eq:Zmn}
{{\tensor{Z} }_{\Delta m,\Delta n}} = j\frac{{{k_0}{Z_0}}}{{MN}}\sum\limits_{pq} {\tensor{T}_{pq}^{\raisemath{-4pt}{*}} {\tensor{G}_{pq}}{\tensor{T}_{pq}}{e^{ - j2\pi \left( {\frac{{p\Delta m}}{M} + \frac{{q\Delta n}}{N}} \right)}}} .
\end{equation}

In summary, (\ref{eq:LTIMOM}) is the MoM matrix equation corresponding to the structure shown in Fig. \ref{fig:lti_struc}. Using Floquet's theorem, the periodicity in space was exploited such that unknowns only had to be placed within a single spatial period. For each observation position ($m\Delta x$, $n\Delta y$) within a spatial period, $\tensor{\eta}_{mn,m'n'}$ represents the overlap integral between the testing function at the observation position, the surface impedance distribution and the basis functions. Since the system is linear and time-invariant (LTI), this term contains no interactions between fields of different frequencies. Meanwhile, the interactions due to the surrounding medium are captured by the matrix $\tensor{Z}_{m-m',n-n'}$. Since the surrounding medium is LTI, this term also does not contain any interactions between fields of different frequencies.

\subsection{Spatially-Discrete Traveling-Wave Modulated Structure} \label{sec:mom_trav_wave}

\begin{figure}[!t]
\centering
\FPset\lenax{0.55}
\FPset\lenay{0.35}
\FPset\lenaz{0.4}
\FPset\lenk{0.35}
\FPset\angkd{75}
\FPmul\angkr\angkd{0.0175}
\FPset\angpd{10}
\FPmul\angpr\angpd{0.0175}
\FPcos\ksx\angkr
\FPsin\ksy\angkr
\FPtan\ptan\angpr
\FPset\lent{.01}
\FPset\rx{-35}
\FPset\ry{35.5}
\FPset\hh{.05}
\FPset\gap{.065}
\FPset\xh{.425}
\FPset\yh{.19}
\FPset\cang{22}
\FPset\dcy{.0455}
\FPset\dcx{.0515}
\FPset\dca{1}
\begin{tikzpicture}
	\definecolor{myred}{rgb}{0.40,0.00,0.60};
	\definecolor{myblue}{rgb}{0,.4,1};
	\definecolor{mygreen}{rgb}{0,.7,.4};
    \node[anchor=south west,inner sep=0](image)at(0,0,0){\includegraphics[clip,trim=0cm 3cm 0cm 3cm,width=.95\columnwidth]{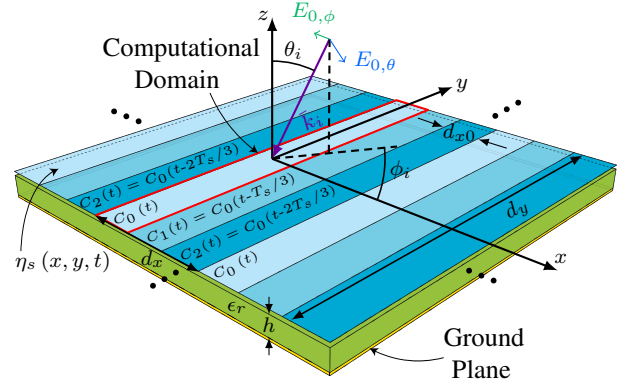}};
    \begin{scope}[x={(image.south east)},y={(image.north west)},font=\fontsize{8}{6}]
        \begin{scope}[shift={(0.43,0.63)}] 
          \draw[black,thick,-latex] coordinate (o) (0,0)  -- coordinate (x) (\rx:\lenax) node[above=4.5pt,right=-5pt] {$x$};
          \draw[black,thick,-latex] (o) -- coordinate[pos=1] (y)(\ry:\lenay) node[above=.5pt,right=-2.5pt] {$y$};
          \draw[black,thick,-latex] (o) -- coordinate[pos=1] (z)(90:\lenaz) node[above=.5pt,left=-2.5pt] {$z$};
          \draw[myred,thick,latex-] (o) -- coordinate[pos=1] (k)(\angkd:\lenk) node[midway,rotate=30,below=9pt,right=-8pt] {$\bar k_i$};
          \draw[black,thick,densely dashed] (k) -- coordinate[pos=1] (p) (\ksx*\lenk,\ksx*\lenk*\ptan);
          \draw[black,thick,densely dashed] (o) -- (\angpd:.2);
          \draw[myblue,->] [shift={(\angkd:\lenk)}] (0,0) -- (-70:.07) node[right] {$E_{0,\theta}$};
          \draw[mygreen,->] [shift={(\angkd:\lenk)}] (0,0) -- (180-35:.03) node[above] {$E_{0,\phi}$};
          \draw pic [draw=black,fill=none,angle radius=13mm,semithick] {angle = k--o--z} node[above=41pt,right=1pt] {$\theta_i$};
          \draw pic [draw=black,fill=none,angle radius=15mm,semithick] {angle = x--o--p} node[below=4pt,right=41pt] {$\phi_i$};
        \end{scope}
        
        \begin{scope}[font=\fontsize{8}{6}]
	        \draw[black,semithick,latex-latex] (.15,.47)  -- (.315,.31) node[midway,below=5.7pt,right=-6pt] {$d_x$};
	        \draw[black,semithick,latex-latex] (.46,.175)  -- (.92,.64) node[near end, sloped,below=-2pt] {$d_{y}$};
	        \draw[black,-latex] (.66,.745)  -- (.7,.72);
	        \draw[black,latex-] (.755,.685)  -- (.8,.66);
	        \node[align=center] (c) at (.726,.704) {$d_{x0}$};
        \end{scope}

		\begin{scope}[shift={(0.345,0.57)}, font=\fontsize{5}{6}]
			\coordinate (o) (0,0);
			\node[align=left,text width=3cm,rotate=\cang-2*\dca] (l) at (-\dcx,.8*\dcy) {$C_2(t)=C_0(t \mhyphen 2T_s/3 )$};
			\node[align=left,text width=3cm,rotate=\cang] (l) at (0,0) {$C_0\left(t\right)$};
			\node[align=left,text width=3cm,rotate=\cang+\dca] (l) at (\dcx,-\dcy) {$C_1(t)=C_0(t \mhyphen T_s/3)$};
			\node[align=left,text width=3cm,rotate=\cang+2*\dca] (l) at (2*\dcx,-2*\dcy) {$C_2(t)=C_0(t \mhyphen 2T_s/3)$};
			\node[align=left,text width=3cm,rotate=\cang+3*\dca] (l) at (3*\dcx,-3*\dcy) {$C_0\left(t\right)$};
		\end{scope}        
        
        \begin{scope}[shift={(\xh,\yh)}]
	        \draw[black,-latex] (0,\hh)  -- (0,0);
	        \draw[black,-latex] (0,-\hh-\gap)  -- (0,-\gap);
	        \node[align=center] (c) at (0,-0.45*\gap) {$h$};
	        \node[align=center] (c) at (-.05,.02) {$\epsilon_r$};
        \end{scope}
		
		\node[align=center,text width=2cm, text=black] (c) at (0.1,.34) {$\mathclap{\eta_s \left(x,y,t\right)}$};
		\node[align=center,text width=1cm, text=black] (c) at (0.76,.1) {{\fontsize{12}{2} Ground Plane}};
		\node[align=center,text width=3cm, text=black] (c) at (0.28,.9) {{\fontsize{12}{2} Computational Domain}};		
		
		\node[align=center,text width=2cm, text=black,rotate=-22] (c) at (.76,0.3) {{\huge ...}}; 
		\node[align=center,text width=2cm, text=black,rotate=-22] (c) at (.2,0.75) {{\huge ...}}; 
		\node[align=center,text width=2cm, text=black,rotate=30] (c) at (.26,0.29) {{\huge ...}}; 
		\node[align=center,text width=2cm, text=black,rotate=30] (c) at (.8,0.77) {{\huge ...}}; 
    \end{scope}
\end{tikzpicture}
\caption{An example of a 3-stixel SD-TWM structure examined in \cite{zca}. Since there are 3 stixels per supercell ($L=3$) the time-dependence of the capacitance in stixel $l$ is given by ${C_l}\left( t \right) = {C_0}\left( {t - {\textstyle{l \over L}}{T_s}} \right) = {C_0}\left( {t - {\textstyle{l \over 3}}{T_s}} \right)$. The width of each stixel is $d_{x0}={d_x}/{L}={d_x}/{3}$. The presented formulation is valid for periodic variations along $y$, although this is not depicted in the figure.}
\label{fig:cap_strips}
\end{figure}

In this section, the MoM formulation for the SD-TWM structure shown in Fig. \ref{fig:cap_strips} will be derived. However, it should be noted that the following procedure can easily extended to simulate other SD-TWM structures. The MoM formulation will be obtained by modifying the time-invariant analysis in Section \ref{sec:mom_LTI} to account for a SD-TWM impedance sheet.
The interpath relation in (\ref{eq:interpath_fd}) will be used to construct basis functions such that unknowns will only need to be placed within a single stixel.
Therefore, the following analysis procedure reduces the number of unknowns by a factor of $L$, the number of stixels in a supercell.

The structure in Fig. \ref{fig:cap_strips} consists of time-varying capacitive strips placed on the surface of a grounded dielectric substrate. The time variation of the capacitance within each stixel is staggered in time. That is, the sheet capacitance in stixel $l$, $C_l\left( t \right)$, satisfies
\begin{equation} \label{eq:cap_relation}
{C_l}\left( t \right) = {C_{l - 1}}\left( {t - {T_s}/L} \right) ,
\end{equation}
where $L$ is the number of stixels in a supercell. The space-time dependent sheet capacitance over the entire supercell can then be written
\begin{equation} \label{eq:cap_super}
C\left( {x,t} \right) = {C_l}\left( t \right),\quad x \in \left[ {\left( l - {\textstyle{1 \over 2}} \right){d_{x0}},\left( {l + {\textstyle{1 \over 2}}} \right){d_{x0}}} \right) .
\end{equation}
The excitation field is the same as in Section \ref{sec:mom_LTI}, and can be written in the time-domain as
\begin{align} \label{eq:excite}
\begin{split}
\bar {\mathcal{E}_i}\left( {\bar r,t} \right) &= {{\bar E}_0}{e^{j\left( {{\omega _0}t - {{\bar k}_i} \cdot \bar r} \right)}} \\
& = {{\bar E}_i}\left( {\bar r} \right){e^{j{\omega _0}t}}
\end{split} .
\end{align}

Similar to the time-invariant formulation, we will first write the boundary condition for the total field at $z=0$. For a sheet capacitance, this boundary condition is given by
\begin{equation} \label{eq:BC_TV1}
\frac{\partial }{{\partial t}}\left\{ {C\left( {x,t} \right){{\bar {\mathcal{E}}}_t}\left( {x,y,t} \right)} \right\} = {\bar {\mathcal{J}}_s}\left( {x,y,t} \right) ,
\end{equation}
where ${{\bar {\mathcal{E}}}_t}\left( {x,y,t} \right)$ denotes the transverse component of the total time-dependent electric field. However, for our formulation, it is simpler to implement the following, equivalent boundary condition
\begin{align} \label{eq:BC_TV2}
\begin{split}
\bar {\mathcal{E}_{t}}\left( {x,y,t} \right) &= \frac{1}{{C\left( {x,t} \right)}}\int {{{\bar {\mathcal{J}}}_s}\left( {x,y,t} \right)dt}  \\
& = \frac{{j{\omega _0}}}{{j{\omega _0}C\left( {x,t} \right)}}\int {{{\bar {\mathcal{J}}}_s}\left( {x,y,t} \right)dt} \\
& = {\eta _s}\left( {x,y,t} \right)j{\omega _0}\int {{{\bar {\mathcal{J}}}_s}\left( {x,y,t} \right)dt} 
\end{split} ,
\end{align}
where ${\eta _s}\left( {x,y,t} \right)  \buildrel \Delta \over =  {1 \mathord{\left/
 {\vphantom {1 {j{\omega _0}C\left( {x,y,t} \right)}}} \right.
 \kern-\nulldelimiterspace} {j{\omega _0}C\left( {x,t} \right)}}$ is the effective time-varying surface impedance. Since ${C\left( {x,t} \right)}$ is a periodic function in time, ${\eta _s}\left( {x,y,t} \right)$ is also periodic, and can be expanded in a Fourier series as
\begin{equation} \label{eq:•}
{\eta _s}\left( {x,y,t} \right) = \sum\limits_{\nu  =  - \infty }^\infty  {\eta _s^\nu \left( {x,y} \right){e^{j \nu {\omega _s}t}}} .
\end{equation}
Further, since the excitation field in (\ref{eq:excite}) has a time-dependence of $e^{j\omega_0 t}$ and the structure is an LPTV system, the electric field and surface current density can be written as \cite{richards}
\begin{align} \label{eq:•}
\begin{split}
{{\bar {\mathcal{E}}}_t}\left( {x,y,t} \right) &= \sum\limits_{\nu  =  - \infty }^\infty  {\bar E_t^\nu \left( {x,y} \right){e^{j\left( {{\omega _0} + \nu {\omega _s}} \right)t}}} \\
{{\bar {\mathcal{J}}}_s}\left( {x,y,t} \right) &= \sum\limits_{\nu ' =  - \infty }^\infty  {\bar J_s^{\nu '}\left( {x,y} \right){e^{j\left( {{\omega _0} + \nu ' {\omega _s}} \right)t}}} 
\end{split} .
\end{align}
Note that this implies
\begin{equation} \label{eq:intJ}
j{\omega _0}\int {{{\bar J}_s}\left( {x,y,t} \right)dt}  = \sum\limits_{\nu ' =  - \infty }^\infty  {\frac{{\bar J_s^{\nu '}\left( {x,y} \right)}}{{1 + \nu '{\textstyle{{{\omega _s}} \over {{\omega _0}}}}}}{e^{j\left( {{\omega _0} + \nu' {\omega _s}} \right)t}}} .
\end{equation}
Substituting this expression into (\ref{eq:BC_TV2}) yields
\begin{equation} \label{eq:BC_TV3}
\bar E_t^\nu \left( {x,y} \right) = \sum\limits_{\nu ' =  - \infty }^\infty  {\frac{{\eta _s^{\nu  - \nu '}\left( {x,y} \right)}}{{1 + \nu '{\textstyle{{{\omega _s}} \over {{\omega _0}}}}}}\bar J_s^{\nu '}\left( {x,y} \right)} .
\end{equation}
This is expected since a product in the time-domain results in convolution in the frequency-domain. Equation (\ref{eq:BC_TV3}) is the boundary condition we will use in the MoM formulation. Note the similarity between this expression and (\ref{eq:BC1}). In effect, (\ref{eq:BC_TV3}) will yield an integral equation for each ``observed'' frequency harmonic, $\nu$, which is coupled to all ``source'' frequency harmonics, $\nu'$, through the surface impedance.
We will now split the total electric field harmonics into incident and scattered components.
The electric field in the absence of the space-time modulated impedance sheet contains both the excitation field as well as the reflected field from the grounded dielectric. In the time-domain, the reflected field is given by
\begin{equation} \label{eq:td_ref}
{{\bar {\mathcal{E}}}_{r,t}}\left( {x,y,t} \right) = \tensor{\Gamma} {{\bar {\mathcal{E}}}_{i,t}}\left( {x,y,t} \right) ,
\end{equation}
where $\tensor{\Gamma}$ is the dyadic reflection coefficient which is a function of $\theta_i$, $\phi_i$, $h$ and $\epsilon_r$. 
Splitting the total field in (\ref{eq:BC_TV3}) into incident and scattered components, we obtain
\begin{align} \label{eq:BC_TV4}
\begin{split}
\left[ {1 + \mathord{\buildrel{\lower3pt\hbox{$\scriptscriptstyle\leftrightarrow$}} 
\over \Gamma } } \right] & {\bar E_{i,t}}\left( {x,y} \right)\delta_{\nu} 
+ \bar{E}_{s,t}^{\nu}\left(x,y\right) \\
&= \sum\limits_{\nu ' =  - \infty }^\infty  {\frac{{\eta _s^{\nu  - \nu '}\left( {x,y} \right)}}{{1 + \nu '{\textstyle{{{\omega _s}} \over {{\omega _0}}}}}}\bar J_s^{\nu '}\left( {x,y} \right)} 
\end{split} ,
\end{align}
where ${\bar E_{i,t}}\left( {x,y} \right)$ is the transverse component of ${\bar E_{i}}\left( \bar r \right)$ in (\ref{eq:excite}) and $\bar{E}_{s,t}^{\nu}\left(x,y\right)$ is the transverse component of the $\nu^{\rm{th}}$ frequency harmonic of the scattered electric field. The Kronecker delta function which multiplies the incident field in (\ref{eq:BC_TV4}) results from the monochromatic excitation.

We will now express the scattered electric field frequency harmonics in terms of the surface current density.
Because the surrounding medium is time-invariant (only the surface impedance is time-dependent), the scattered field at frequency $f_0+\nu f_s$ only depends on the induced current at the same frequency.
Since the supercell is periodic in $x$ and $y$, it follows from Floquet's theorem that
\begin{equation} \label{eq:Jexp1}
\bar J_s^{\nu }\left( {x,y} \right) = {e^{ - j\left( {k_{ix}x + {k_{iy}}y} \right)}}\bar j_s^{\nu }\left( {x,y} \right) ,
\end{equation}
where $\bar j_s^{\nu }\left( {x,y} \right)$ is a periodic function in $x$ with period $d_x$ and in $y$ with period $d_y$. Since $\bar j_s^{\nu }\left( {x,y} \right)$ is periodic in $x$ and $y$, it can be expanded in terms of 2D Fourier series as
\begin{equation} \label{eq:•}
{\bar j^{\nu}_s}\left( {x,y} \right) = \sum\limits_{p =  - \infty }^\infty  {\sum\limits_{q =  - \infty }^\infty  {{{\bar I}^{\nu }_{pq}}{e^{-j\left( {{k'_{xp}}x + {k'_{yq}}y} \right)}}} } ,
\end{equation}
where ${k'_{xp}} = {\textstyle{{2\pi p} \over d_{x}}}$ and ${k'_{yq}} = {\textstyle{{2\pi q} \over d_y}}$.
Substituting this into (\ref{eq:Jexp1}), we obtain
\begin{equation} \label{eq:Jexp2}
\bar J_s^{\nu }\left( {x,y} \right) = \sum\limits_{p =  - \infty }^\infty  {\sum\limits_{q =  - \infty }^\infty  {{{\bar I}^{\nu }_{pq}}{e^{ - j\left( {k_{xp}x + {k_{yq}}y} \right)}}} } ,
\end{equation}
where $k_{xp} = k_{ix} + k'_{xp}$ and $k_{yq} = k_{iy}+k'_{yq}$.
Just as in Section \ref{sec:mom_LTI}, we can interpret (\ref{eq:Jexp2}) as a superposition of planar current sheets of the form ${\bar J_0}{e^{ - j\left( {{k_x}x + {k_y}y} \right)}}$. Therefore, the electric field scattered by the surface current density can be written as \cite{jin}
\begin{multline} \label{eq:escat_tv}
{\bar E^{\nu}_{s,t}}\left( {x,y} \right) = \\ 
 - j{k_0}{Z_0}\sum\limits_{p  - \infty }^\infty  {\sum\limits_{q =  - \infty }^\infty  {\tensor{G}\cup{\nu}\left( {{k_{xp}},{k_{yq}}} \right){{\bar I}^{\nu}_{pq}}{e^{ - j\left( {{k_{xp}}x + {k_{yq}}y} \right)}}} }   ,
\end{multline}
where $\tensor{G}\cup{\nu}\left( {{k_{xp}},{k_{yq}}} \right)$ is the spectral-domain representation of the dyadic Green's function as derived in Appendix \ref{sec:A1} evaluated at frequency $f_0+\nu f_s$. Substituting the expansion for the scattered electric field into (\ref{eq:BC_TV4}) and multiplying both sides by $e^{j(k_{ix}x + {k_{iy}}y)}$, we obtain
\begin{multline} \label{eq:integral_eqn_tv}
\left[ {1 + \tensor{\Gamma} } \right]  
{\bar E_{0,t}} \delta _\nu = 
 \sum\limits_{\nu ' =  - \infty }^\infty  {\frac{{\eta _s^{\nu  - \nu '}\left( {x,y} \right)}}{{1 + \nu '{\textstyle{{{\omega _s}} \over {{\omega _0}}}}}} \bar j_s^{\nu '}\left( {x,y} \right)} \quad  \\
+ j k_0 Z_0 \sum\limits_{p =  - \infty }^\infty  {\sum\limits_{q =  - \infty }^\infty  {\Gpqn{{\bar I}^{\nu}_{pq}}{e^{ - j\left( {k'_{xp} x + {k'_{yq}}y} \right)}}} } ,
\end{multline}
where $\bar E_{0,t}$ is the transverse component of the excitation field amplitude and $\Gpqn=\tensor{G}\cup{\nu}\left( {{k_{xp}},{k_{yq}}} \right)$. Equation (\ref{eq:integral_eqn_tv}) represents our integral equation since $\bar I^{\nu}_{pq}$ is computed via a spatial integral over the supercell
\begin{equation} \label{eq:2dft_tv}
{{\bar I}^{\nu}_{pq}} = \frac{1}{{d_{x} d_y}}\int\limits_{-\frac{d_{x0}}{2}}^{d_x-\frac{d_{x0}}{2}} {\int\limits_{ - \frac{d_{y}}{2}}^{\frac{d_{y}}{2}} {{{\bar j}^{\nu}_s}\left( {x,y} \right){e^{j\left( {{k'_{xp}}x + {k'_{yq}}y} \right)}}dxdy} } .
\end{equation}


\FPset\lenax{0.2}
\FPset\lenay{0.185}
\FPset\lenaz{0.25}
\FPset\lent{.01}
\FPset\rx{-35}
\FPset\ry{32}
\begin{figure}[!t]
\centering
\begin{tikzpicture}
    \node[anchor=south west,inner sep=0](image)at(0,0,0){\includegraphics[clip,trim=0cm 3cm 0cm 3cm, width=.95\columnwidth]{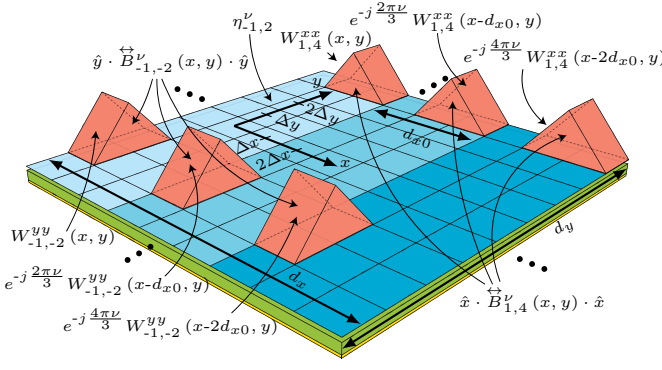}};
    \begin{scope}[x={(image.south east)},y={(image.north west)},font=\fontsize{6}{6}]
        \begin{scope}[shift={(0.35,0.66)}] 
          \draw[black,thin,shift={(\rx:0.06)}] (180+\ry:\lent) -- (\ry:\lent) node[rotate=25,below=-1pt,left=1pt] {$\Delta x$};
          \draw[black,thin,shift={(\rx:2*0.06)}] (180+\ry:\lent) -- (\ry:\lent) node[rotate=25,below=-1pt,left=1pt] {$2\Delta x$};
          \draw[black,thin,shift={(\ry:0.06)}] (180+\rx:\lent) -- (\rx:\lent) node[rotate=-20,above=0pt,right=-2pt] {$\Delta y$};
          \draw[black,thin,shift={(\ry:2*0.06)}] (180+\rx:\lent) -- (\rx:\lent) node[rotate=-20,above=-.9pt,right=-2.7pt] {$2\Delta y$};
          \draw[black,thick,-latex] coordinate (o) (0,0)  -- coordinate (a) (\rx:\lenax) node[below=-1pt,right=-3pt] {$x$};
          \draw[black,thick,-latex] (o) -- coordinate (c)(\ry:\lenay) node[above=3pt,right=-12pt] {$y$};
        \end{scope}
        \draw[black,thick,latex-latex] (.057,.565)  -- (.55,.1) node[near end,sloped,above=4pt,right=-3pt] {$d_x$};
        \draw[black,thick,latex-latex] (.57,.705)  -- (.725,.62) node[midway, sloped,below=5pt,right=-10pt] {$d_{x0}$};
        \draw[black,thick,latex-latex] (.52,.03)  -- (.97,.54) node[near end, sloped,below=-1pt] {$d_{y}$};
		
		\node[align=center,text width=2cm, text=black] (c) at (0.82,0.17) {$\mathclap{\hat{x}\cdot\tensor{B}_{1,4}\cup{\nu}\left(x,y\right)\cdot\hat{x}}$};
		\node[align=center,text width=2cm, text=black] (c) at (0.495,.905) {$\mathclap{W_{1,4}\cup{xx}\left(x,y\right)}$};
		\node[align=center,text width=2cm, text=black] (c) at (0.65,.965) {$e^{\mhyphen j\frac{2\pi\nu}{3}}W_{1,4}\cup{xx}\left(x \mhyphen d_{x0},y\right)$};
		\node[align=center,text width=2cm, text=black] (c) at (0.83,.86) {$e^{\mhyphen j\frac{4\pi\nu}{3}}W_{1,4}\cup{xx}\left(x \mhyphen 2d_{x0},y\right)$};
		
		\node[align=center,text width=2cm, text=black] (c) at (0.25,0.85) {$\mathclap{\hat{y}\cdot\tensor{B}_{\mhyphen 1,\mhyphen 2}\cup{\nu}\left(x,y\right)\cdot\hat{y}}$};
		\node[align=center,text width=2cm, text=black] (c) at (0.08,0.35) {$\mathclap{W_{\mhyphen 1,\mhyphen 2}\cup{yy}\left(x,y\right)}$};
		\node[align=center,text width=2cm, text=black] (c) at (0.15,0.22) {$\mathclap{e^{\mhyphen j\frac{2\pi\nu}{3}}W_{\mhyphen 1,\mhyphen 2}\cup{yy}\left(x \mhyphen d_{x0},y\right)}$};
		\node[align=center,text width=2cm, text=black] (c) at (0.25,0.1) {$\mathclap{e^{\mhyphen j\frac{4\pi\nu}{3}}W_{\mhyphen 1,\mhyphen 2}\cup{yy}\left(x \mhyphen 2d_{x0},y\right)}$};
		\node[align=center,text width=2cm, text=black] (c) at (.38,0.95) {$\mathclap{\eta_{\mhyphen 1,2}^\nu}$};
		\node[align=center,text width=2cm, text=black,rotate=-22] (c) at (.82,0.27) {{\huge ...}};
		\node[align=center,text width=2cm, text=black,rotate=-22] (c) at (.28,0.75) {{\huge ...}};
		\node[align=center,text width=2cm, text=black,rotate=30] (c) at (.2,0.3) {{\huge ...}};
		\node[align=center,text width=2cm, text=black,rotate=30] (c) at (.665,0.775) {{\huge ...}};
    \end{scope}
\end{tikzpicture}
\caption{An overlay of the basis functions used to expand the current within a supercell of the structure shown in Fig. \ref{fig:cap_strips}. In this example, there are 3 stixels within a supercell ($d_x=3d_{x0}$). There are only $2M'+1=3$ unique unknowns along $x$ since the basis functions account for the interpath relation, reducing the number of unknowns with respect to Fig. \ref{fig:basis}. There are $2N'+1=9$ unknowns along $y$}
\label{fig:basis_tv}
\end{figure}

Numerically computing the current requires us to expand ${{\bar j}^{\nu}_s}\left( {x,y} \right)$ into a set of basis functions. It should be noted that, up to this point, we have not invoked the symmetry of the SD-TWM structure. In fact, using the same basis functions as those described in Section \ref{sec:mom_LTI} would result in the standard harmonic-balance MoM approach to solving the structure in Fig. \ref{fig:cap_strips}. However, from the interpath relation in (\ref{eq:interpath_fd}), it is clear that the fields in a single stixel should be sufficient to solve the entire problem.
Thus, to reduce the number of unknowns required to simulate the structure, we will modify the basis functions applied to the time-invariant problem in Section \ref{sec:mom_LTI}. The modified basis will be constructed such that:
\begin{itemize}
\item surface current weighting coefficients only need to be placed within a single stixel
\item the modified basis functions satisfy the interpath relation in (\ref{eq:interpath_fd})
\end{itemize}
We will define $\tensor{W}_{mn}\left(x,y\right)$ to be the same as in (\ref{eq:Wxx}) through (\ref{eq:lamy}). However, in this section, we require the computational grid to be identical within each stixel. Therefore, we redefine $M$ as the number of computational elements within a stixel (rather than a spatial period). This re-scales the width of the computational elements to be $\Delta x = d_{x0}/M$ rather than $d_x / M$.
We will now construct a new basis function $\tensor{B}\cup{\nu}_{m'n'}\left( {x,y} \right)$, shown in Fig. \ref{fig:basis_tv}, which is given by
\begin{equation} \label{eq:basis_tv}
{\tensor{B}\cup{\nu}_{m'n'}}\left( {x,y} \right) = \sum\limits_{l = 0}^{L - 1} {{e^{ - j{\textstyle{{2\pi \nu l} \over L}}}}{\tensor{W}_{m'n'}}\left( {x - l{d_{x0}},y} \right)} .
\end{equation}
Note that $\tensor{B}\cup{\nu}_{m'n'} \left(x,y\right)$ satisfies the interpath relation in (\ref{eq:interpath_fd}) and spans the entire supercell.
The expansion of ${{\bar j}^{\nu}_s}\left( {x,y} \right)$ in terms of basis functions $\tensor{B}\cup{\nu}_{m'n'} \left(x,y\right)$ is given by
\begin{equation} \label{eq:basis_exp_tv}
\bar j_s^\nu \left( {x,y} \right) = \sum\limits_{n =  - N'}^{N'} {\sum\limits_{m =  - M'}^{M'} \tensor{B}\cup{\nu}_{m'n'}\left(x,y\right)  \bar j_{m'n'}^\nu } .
\end{equation}
We can substitute this expansion into (\ref{eq:2dft_tv}) to obtain $\bar I^{\nu}_{pq}$ in terms of the weighting coefficients ${{\bar j}^{\nu}_{m'n'}}$, which yields
\begin{multline} \label{eq:Ipq_tv1}
{{\bar I}^{\nu}_{pq}} = \\
\frac{\tensor{T}_{pq}}{{LMN}}\sum\limits_{m'n'} {
\sum\limits_{l = 0}^{L - 1} {{e^{j{\textstyle{{2\pi \left( {p - \nu } \right)l} \over L}}}}} 
{{\bar j}^{\nu}_{m'n'}}{e^{j\left( \frac{2\pi pm'}{LM} + \frac{2\pi qn'}{N} \right)}}} ,
\end{multline}
where ${\tensor{T}_{pq}}$ represents the 2D Fourier series coefficients of ${\tensor{W}_{00}}$ given by
\begin{align} \label{eq:Tpq_tv}
\begin{split}
\tensor{T}_{pq}  = & {\rm sinc} ^2 \left( \frac{p\pi}{LM} \right) {\rm sinc} \left( \frac{q\pi}{N} \right)e^{j\frac{p\pi}{LM}} \hat x \hat x 
\\ + & {\rm sinc} \left( \frac{p\pi}{LM} \right) {\rm sinc} ^2 \left( \frac{q\pi}{N} \right)e^{j\frac{q\pi}{N}} \hat y \hat y 
\end{split} .
\end{align}
The factor of $1/L$ in (\ref{eq:Ipq_tv1}) and (\ref{eq:Tpq_tv}) results from an effective increase in discretization by a factor of $L$.
This expression can be further simplified by noting
\begin{equation} \label{eq:•}
\sum\limits_{l = 0}^{L - 1} {{e^{j{\textstyle{{2\pi \left( {p - \nu } \right)l} \over L}}}}}  = \left\{ {\begin{array}{*{20}{l}}
{L,\quad p - \nu  = p'L,p' \in \mathbb{Z}}\\
{0,\quad {\rm{otherwise}}}
\end{array}} \right. .
\end{equation}
If we define
\begin{equation} \label{eq:g}
g_p^\nu  = \left\{ {\begin{array}{*{20}{l}}
{1,\quad p - \nu  = p'L,p' \in \mathbb{Z}}\\
{0,\quad {\rm{otherwise}}}
\end{array}} \right. ,
\end{equation}
then the summation over $l$ (the stixels) in (\ref{eq:Ipq_tv1}) can be eliminated. Thus (\ref{eq:Ipq_tv1}) can be written as
\begin{equation} \label{eq:Ipq_tv2}
{{\bar I}^{\nu}_{pq}} = \frac{\tensor{T}_{pq}}{{MN}}\sum\limits_{m'n'} {g_p^\nu
{{\bar j}^{\nu}_{m'n'}}{e^{j\left( {2\pi pm'/LM + 2\pi qn'/N} \right)}}}  .
\end{equation}
We now substitute $\bar j_s^\nu$ from (\ref{eq:basis_exp_tv}) and ${\bar I}^{\nu}_{pq}$ from (\ref{eq:Ipq_tv2}) into (\ref{eq:integral_eqn_tv}) to obtain an expression relating the incident field amplitude to the unknown current weighting coefficients.
\begin{multline} \label{eq:pretest_tv}
\left[ {1 + \tensor{\Gamma} } \right]{{\bar E}_{0,t}}{\delta _\nu } = 
\sum\limits_{\nu '} {\sum\limits_{m'n'} {\frac{{\eta _s^{\nu  - \nu '}\left( {x,y} \right)}}{{1 + \nu '{\textstyle{{{\omega _s}} \over {{\omega _0}}}}}}
{\tensor{B}\cup{\nu}_{m'n'}}\left( {x,y} \right)\bar j_{m'n'}^{\nu'} } }  \\
+ j\frac{{{k_0}{Z_0}}}{{MN}}\sum\limits_{m'n'} {\sum\limits_{pq}
{{e^{ - j\left( {{k'_{xp}}x + {k'_{yq}}y} \right)}}g_p^\nu\Gpqn{\tensor{T}_{pq}}H_{m'n'}^{pq}\bar j_{m'n'}^\nu } } ,
\end{multline}
where 
$H_{m'n'}^{pq}={{e^{j\left( {2\pi pm'/LM + 2\pi qn'/N} \right)}}}$ is the phase term from (\ref{eq:Ipq_tv2}).
It is worth noting the similarity between this expression and (\ref{eq:pretest}). The first term on the right-hand side of (\ref{eq:pretest_tv}) now includes a summation over source frequencies $f_0+\nu ' f_s$, representing the coupling between frequencies introduced by the time-varying impedance sheet. Meanwhile, the only difference in the second term of these two equations is the factor $g_p^\nu$ which captures the induced tangential momentum imparted by the SD-TWM.

The final MoM matrix equation can be obtained by testing the left- and right-hand sides of (\ref{eq:pretest_tv}) for observation points within a single stixel
\begin{multline} \label{eq:•}
\frac{1}{{d_{x} d_y}}\int\limits_{ - \frac{d_{x0}}{2}}^{d_x-\frac{d_{x0}}{2}} {\int\limits_{ - \frac{d_{y}}{2}}^{\frac{d_{y}}{2}} {{\tensor{W}_{mn}}\left( {x,y} \right)\left\{  \bullet  \right\}dxdy} } \\
\forall \left\{ {m \in \left[ { - M',M'} \right],n \in \left[ { - N',N'} \right]} \right\}.
\end{multline}
Note that, while the integral bounds span the entire supercell, the observation positions $(m\Delta x,n \Delta y)$ are limited to a single stixel.
For this implementation, we will approximate the surface impedance frequency harmonics as summations over pulse functions.
\begin{equation} \label{eq:Zs_tv}
\eta_s^{\nu}\left( {x,y} \right) = \sum\limits_{m'' =  - M'}^{M'} {\sum\limits_{n'' =  - N'}^{N'} {{\Pi _{m''}}\left( x \right){\Pi _{n''}}\left( y \right)\eta^{\nu}_{m''n''}} } 
\end{equation}
Carrying out the integrations, we obtain the final MoM matrix equation, which can be written as
\begin{multline} \label{eq:TVMOM}
\left[ {1 + \tensor{\Gamma} } \right]{{\bar E}_{0,t}}{\delta _\nu } = \\
\sum\limits_{\nu '} {\sum\limits_{m'n'} {\left( {
\frac{\tensor{\eta}_{mn,m'n'}^{\raisemath{-4pt}{\nu-\nu'}}}{1+\nu'\frac{\omega_s}{\omega_0}}
+ {\tensor{Z}^{\raisemath{-4pt}{\nu'}}_{m - m',n - n'}}{\delta _{\nu -\nu '}}} \right)\bar j_{m'n'}^{\nu '}} } \\
\forall \left\{ {m \in \left[ { - M',M'} \right],n \in \left[ { - N',N'} \right]} \right\} , \qquad
\end{multline}
where
\begin{align} \label{eq:Zxx_TV}
\begin{split}
\hat x \cdot & \tensor{\eta}\cup{\nu}_{mn,m'n'} \cdot \hat x  = \\
& \quad {\delta _{n-n'}}\left\{ {\rar{1.2} \begin{array}{*{20}{l}}
{{\textstyle{1 \over 6}}\eta^\nu_{m - 1,n},\quad m' = m - 1}\\
{{\textstyle{1 \over 3}}\left\{ {\eta^\nu_{m - 1,n} + \eta^\nu_{m,n}} \right\},\quad m' = m}\\
{{\textstyle{1 \over 6}}\eta^\nu_{m,n},\quad m' = m + 1}\\
{0,\quad {\rm{otherwise}}}
\end{array}} \right.
\end{split}
\end{align}
\begin{align} \label{eq:Zyy_TV}
\begin{split}
\hat y \cdot & \tensor{\eta}\cup{\nu} _{mn,m'n'} \cdot \hat y = \\
& \quad {\delta _{m-m'}}\left\{ {\rar{1.2} \begin{array}{*{20}{l}}
{{\textstyle{1 \over 6}}\eta^\nu_{m,n - 1},\quad n' = n - 1}\\
{{\textstyle{1 \over 3}}\left\{ {\eta^\nu_{m,n - 1} + \eta^\nu_{m,n}} \right\},\quad n' = n}\\
{{\textstyle{1 \over 6}}\eta^\nu_{m,n},\quad n' = n + 1}\\
{0,\quad {\rm{otherwise}}}
\end{array}} \right.
\end{split}
\end{align}
\begin{equation} \label{eq:Zxy_TV}
\hat x \cdot \tensor{\eta}\cup{\nu}_{mn,m'n'} \cdot \hat y = \hat y \cdot \tensor{\eta}\cup{\nu}_{mn,m'n'} \cdot \hat x = 0 ,
\end{equation}
and
\begin{align} \label{eq:Zmn_TV}
\begin{split}
{{\tensor{Z} }\cup{\nu} _{\Delta m,\Delta n}} & = j\frac{{{k_0}{Z_0}}}{{MN}}\sum\limits_{pq} {g_p^\nu \tensor{T}_{pq}\cup{*} \Gpqn \tensor{T}_{pq} {e^{ \mhyphen j2\pi \left( {\frac{{p\Delta m}}{LM} + \frac{{q\Delta n}}{N}} \right)}}} \\
& = j\frac{{{k_0}{Z_0}}}{{MN}}{e^{\mhyphen j\frac{{2\pi \nu \Delta m}}{LM}}}\sum\limits_{p'q} {\tensor{F}_{p'q}\cup{\nu} {e^{ \mhyphen j2\pi \left( {\frac{{p'\Delta m}}{M} + \frac{{q\Delta n}}{N}} \right)}}} 
\end{split} \\
\tensor{F}_{p'q}\cup{\nu} &= \tensor{T}_{p'L+\nu_e,q}\cup{*} \tensor{G}\cup{\nu}_{p'L+\nu_e,q} \tensor{T}_{p'L+\nu_e,q} \\
\nu_e&=\rm{mod}\left(\nu,L\right) .
\end{align}

In summary, (\ref{eq:TVMOM}) represents the MoM matrix equation corresponding to the structure shown in Fig. \ref{fig:cap_strips}. Due to the interpath relation, unknowns only need to be placed within a single stixel (as opposed to the entire supercell). For each observed frequency $\nu$ and position ($m\Delta x$, $n\Delta y$) within a single stixel, $\tensor{\eta}\cup{\nu-\nu'}_{mn,m'n'}$ represents the overlap integral between the testing function at the observation position, the spatial distribution of frequency harmonic $\nu-\nu'$ of the effective time-varying surface impedance, and the basis functions corresponding to frequency $\nu'$. Since the surface impedance is LPTV, this term captures the interactions between fields of different frequencies. Meanwhile, the interactions due to the surrounding medium are captured by the matrix $\tensor{Z}^{\mm{-5}{\nu'}}_{\mm{4}{m-m',n-n'}}$. Since the surrounding medium is LTI, this term is only included when the observed frequency is equal to the source frequency (as denoted by $\delta_{\nu- \nu'}$).

The interpath relation serves to reduce the required number of unknowns in the MoM matrix equation. It is clear that, without invoking the interpath relation, the number of unknowns would have scaled as $\mathcal{O}(UNLM)$, where
\begin{itemize}
    \item $U$ - number of simulated frequency harmonics
    \item $N$ - number of unknowns along $y$ within a spatial period
    \item $L$ - number of stixels per supercell
    \item $M$ - number of unknowns along $x$ within a single stixel.
\end{itemize}
By including the interpath relation, the number of unknowns scales as $\mathcal{O}(UNM)$, reducing the problem size by a factor of $L$. As discussed in \cite{zca}, structures built to mimic continuous traveling-wave modulation require an appreciable number of stixels per spatial modulation period. In these scenarios, the reduction in unknowns which results from shrinking the computational domain to a single stixel would be particularly significant.

\section{Numerical Simulation Results} \label{sec:results}

\rar{2}
\begin{table}[]
\centering
\caption{Summary of modulation examples studied in Section \ref{sec:results}.}
\label{tab:params}
\begin{tabular}{|c|c|c|c|}
\hline
\multirow{2}{*}{} & \multicolumn{3}{c|}{Case} \\ \cline{2-4} 
 & \textbf{A} & \textbf{B} & \textbf{C} \\ \hline
\rar{1.2} \begin{tabular}[c]{@{}c@{}}RF Carrier\\ Frequency ($f_0$)\end{tabular} & 10 GHz & 10 GHz & 10 GHz \\ \hline
Incident Angle ($\theta_i$) & 25$^\circ$ & 25$^\circ$ & 25$^\circ$ \\ \hline
\rar{1.2} \begin{tabular}[c]{@{}c@{}}Modulation\\ Frequency ($f_s$)\end{tabular} & 25 kHz & 25 kHz & 500 MHz \\ \hline
Stixel Width ($d_{x0}$) & $\lambda_0 / 5$ & $\lambda_0 / 5$ & $\lambda_0 / 10$ \\ \hline
\rar{1.2} \begin{tabular}[c]{@{}c@{}}Dielectric\\ Thickness ($h$)\end{tabular} & 0.508 mm & 0.508 mm & 0.508 mm \\ \hline
\rar{1.2} \begin{tabular}[c]{@{}c@{}}Stixels per\\ Supercell ($L$)\end{tabular} & 20 & 3 & 60 \\ \hline
\rar{1.2} \begin{tabular}[c]{@{}c@{}}Capacitance\\ Waveform\end{tabular} & sawtooth phase & sawtooth phase & sinusoidal \\ \hline
\end{tabular}
\end{table}

In this section, numerical results of the MoM formulation presented in Section \ref{sec:mom_trav_wave} for the structure shown in Fig. \ref{fig:cap_strips} are validated and discussed. In \cite{zca}, various capabilities are achieved by considering the metasurface in Fig. \ref{fig:cap_strips} as a space-time reflection phase modulator. Consider a uniform sheet capacitance, $C$, placed on top of a grounded dielectric.  For a given polarization and incidence angle, it can be shown that the reflection phase, $\phi$, satisfies \cite{zca}
\begin{equation} \label{eq:phi}
\tan \frac{{{\phi ^{{\rm{TX}}}}}}{2} =  - Z_0^{{\rm{TX}}}{\omega _0}C_0^{\rm{TX}}\Delta_C   ,
\end{equation}
where the superscript TX refers to either TE$_z$ or TM$_z$ polarization, $Z_0^{\rm{TX}}$ is the tangential wave impedance in free space,  $\omega_0$ is RF frequency, $C_0^{\rm{TX}}$ is the resonant value of capacitance for each polarization, and $\Delta_C$ is defined such that $C=C_0^{\rm{TX}}(1+\Delta_C)$. Therefore, given a desired phase variation, (\ref{eq:phi}) can be used to find the required capacitance variation.

Three modulation examples of the structure shown in Fig. \ref{fig:cap_strips} are examined within this section.
Note that, for all examples, the capacitance is assumed to be spatially uniform over each stixel. Further, $\bar k_i$ is assumed to be in the $x$-$z$ plane ($k_{iy}=0$). The problem descriptions for each of the three modulations are summarized in Table \ref{tab:params}.
In Section \ref{sec:results_valid}, the time-variation of the capacitance is designed to produce a sawtooth reflection phase. There are 20 stixels of width $\lambda_0 / 5$ per supercell. The convergence of the MoM formulation is examined, and the results are validated using a spectral-domain method \cite{zca}. In Section \ref{sec:results_subharmmix}, the capacitance variation again produces a sawtooth reflection phase. However, in this case, each supercell contains only 3 stixels (becoming subwavelength) and subharmonic mixing can be observed.
Finally, in Section \ref{sec:results_sin}, a sinusoidal capacitance is applied such that the reflected field contains no power at the fundamental frequency or in the spectral direction. In this example, each supercell contains 60 stixels of width $\lambda_0 / 10$, highlighting the capability of the MoM formulation to efficiently simulate continuous designs.

To examine the accuracy and convergence of the proposed method, we will define the error energy between two given solutions in the spectral-domain. From the interpath relation, it can be shown that the fields corresponding to a SD-TWM structure are written in terms of a compressed double Floquet expansion \cite{zca}. Assuming that there is no variation along $y$, the transverse component of the scattered electric field at $z=0$ can be expanded as \cite{zca}
\begin{multline} \label{eq:e_exp}
{\bar{\mathcal{E}}_{s,t}}\left( {x,y,t} \right) = \\
{e^{j\left( {{\omega _0}t - {k_{ix}}x} \right)}}\sum\limits_{p' =  - \infty }^\infty  {\sum\limits_{\nu  =  - \infty }^\infty  {\bar E_{p'}^\nu {e^{j\nu \left( {{\omega _s}t - \frac{{2\pi }}{{{d_x}}}x} \right)}}{e^{ - j\frac{{2\pi p'}}{{{d_{x0}}}}x}}} }  .
\end{multline}
This is the expansion employed in the spectral-domain method reported in \cite{zca} that will be used to determine the accuracy of the MoM formulation.
From the weighting coefficients, $\bar{j}_{m'n'}^\nu$, computed by the MoM solver, the spectral expansion of the field can be obtained in this form using (\ref{eq:Ipq_tv2}) and (\ref{eq:escat_tv}). Thus, given two solutions for the spectral-domain electric field coefficients, $\bar{E}_{p'}^{\nu (1)}$ and $\bar{E}_{p'}^{\nu (2)}$, we define the error energy as
\begin{equation} \label{eq:err}
\Delta \left( {{{\bar E}^{\left( 1 \right)}},{{\bar E}^{\left( 2 \right)}}} \right) = \sqrt {\frac{{\sum\limits_{\nu  =  - U'}^{U'} {\sum\limits_{p' =  - P'}^{P'} {{{\left| {\bar E_{p'}^{\nu \left( 1 \right)} - \bar E_{p'}^{\nu \left( 2 \right)}} \right|}^2}} } }}{{\sum\limits_{\nu  =  - U'}^{U'} {\sum\limits_{p' =  - P'}^{P'} {{{\left| {\bar E_{p'}^{\nu \left( 2 \right)}} \right|}^2}} } }}} .
\end{equation}
Throughout this section, the summations in (\ref{eq:err}) are truncated such that $U'=P'=30$.

\subsection{Validation and Convergence Study}\label{sec:results_valid}

	\begin{figure}
	\centering
	\includegraphics[clip,trim=.3cm .2cm 1.3cm .2cm,width=\columnwidth]{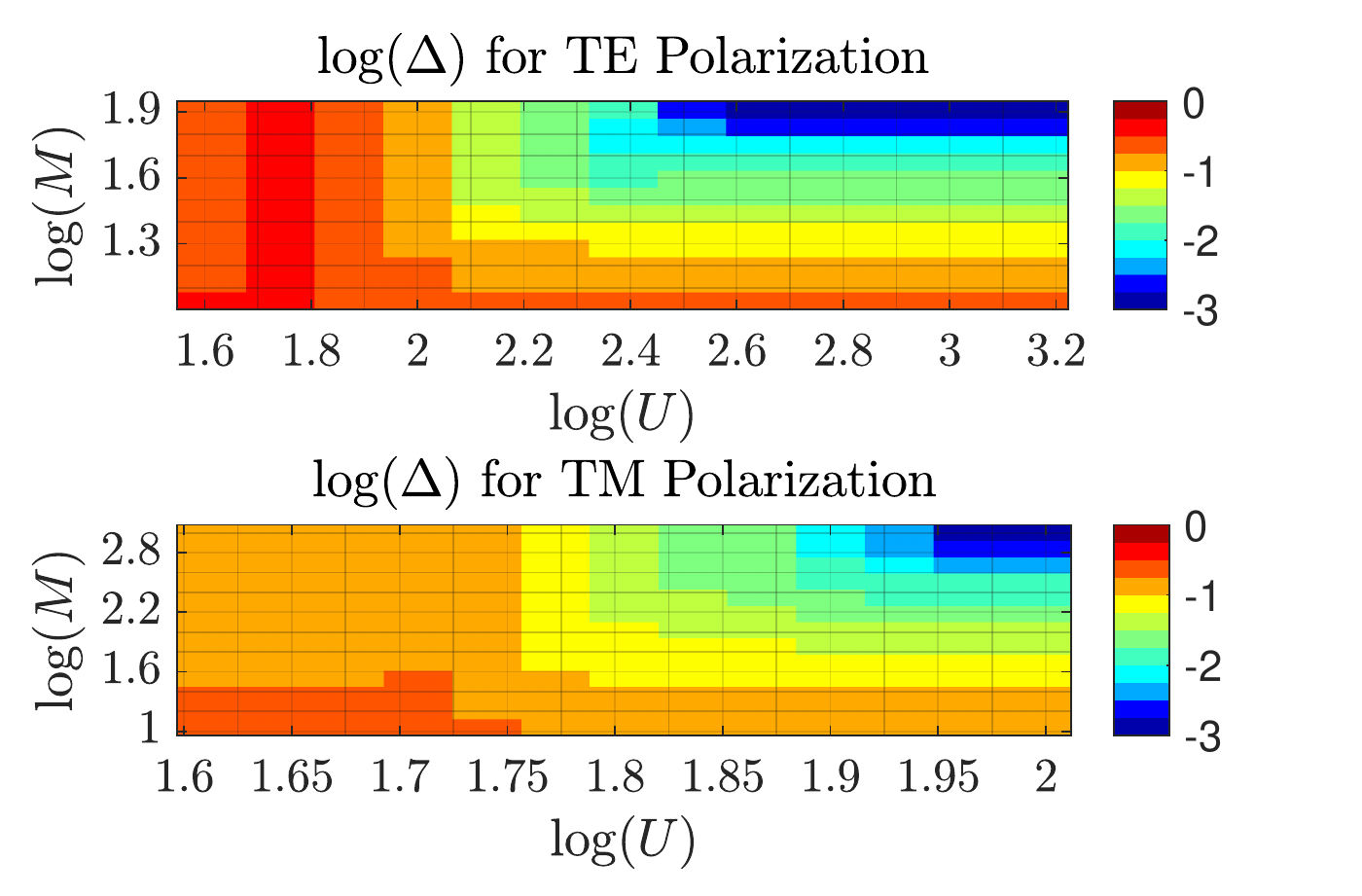}
	\caption{Error energy defined in (\ref{eq:err}) in the electric field harmonic coefficients computed by the MoM solver compared to the finest solution. The number of frequency harmonics is denoted by $U$ while the number of spatial samples is denoted by $M$.}
	\label{fig:converge_mom}
	\end{figure}
	
Here, the convergence and accuracy of the solution obtained by the MoM formulation is examined when the metasurface is modulated as described in case A of Table \ref{tab:params}. The size of each supercell is given by $d_x=Ld_{x0}=4\lambda_0$. In this limit, the sawtooth phase modulation waveform results in a Blaze grating \cite{koshkin1999lower} in both space and time. As a result, we expect the reflected power to be up-converted in frequency and deflected away from the specular direction. The convergence is studied by computing the error energy in (\ref{eq:err}) between the solution for a given number of spatial/spectral unknowns and the solution using the most spatial/spectral unknowns. The accuracy of the formulation is determined by the error energy between the spectral-domain method reported in \cite{zca} and the presented MoM formulation using the most spatial/spectral unknowns.

For a given number of spatial unknowns, $M$, and spectral unknowns, $U$, we define $\bar E ^{(M,U)}$ as the spectral coefficients obtained from the MoM solver. We can then define the error energy in (\ref{eq:err}) with respect to discretization as $\Delta(\bar E ^{(M,U)},\bar E ^{(M_0,U_0)})$, where $M_0$ and $U_0$ are the maximum simulated values of $M$ and $U$ in the convergence study. The solutions for TE$_z$ and TM$_z$ illumination require different numbers of spatial and spectral unknowns to converge. For the TE$_z$ convergence study, $M_0=81$ and $U_0=1,441$. Meanwhile, for the TM$_z$ convergence study, $M_0=999$ and $U_0=99$. In Fig. \ref{fig:converge_mom}, the error energy is shown as a function of $M$ and $U$ for both polarizations. For TE$_z$ polarization, $M=55$ spatial samples and $U=323$ frequency harmonics were required to reduce the error energy below $0.01$. Meanwhile, for TM$_z$ polarization, $M=465$ spatial samples and $U=79$ frequency harmonics were required to reduce the error energy below $0.01$. 
The large number of required unknowns highlights the importance of using the interpath relation to solve SD-TWM systems. For example, the total number of unknowns needed for the TE$_z$ case is $2\times55\times323 \approx 36 \times 10^3$. Without including the interpath relation, the total number of unknowns would have been $20\times2\times55\times323 \approx 711 \times 10^3$.
Despite taking advantage of the inherent sparsity of the matrix system, it would require about 36 GB just to store such a matrix with floating point accuracy. Meanwhile, the interpath relation reduces this requirement to 0.61 GB.

	\begin{figure}
	\centering
	\includegraphics[width=\columnwidth]{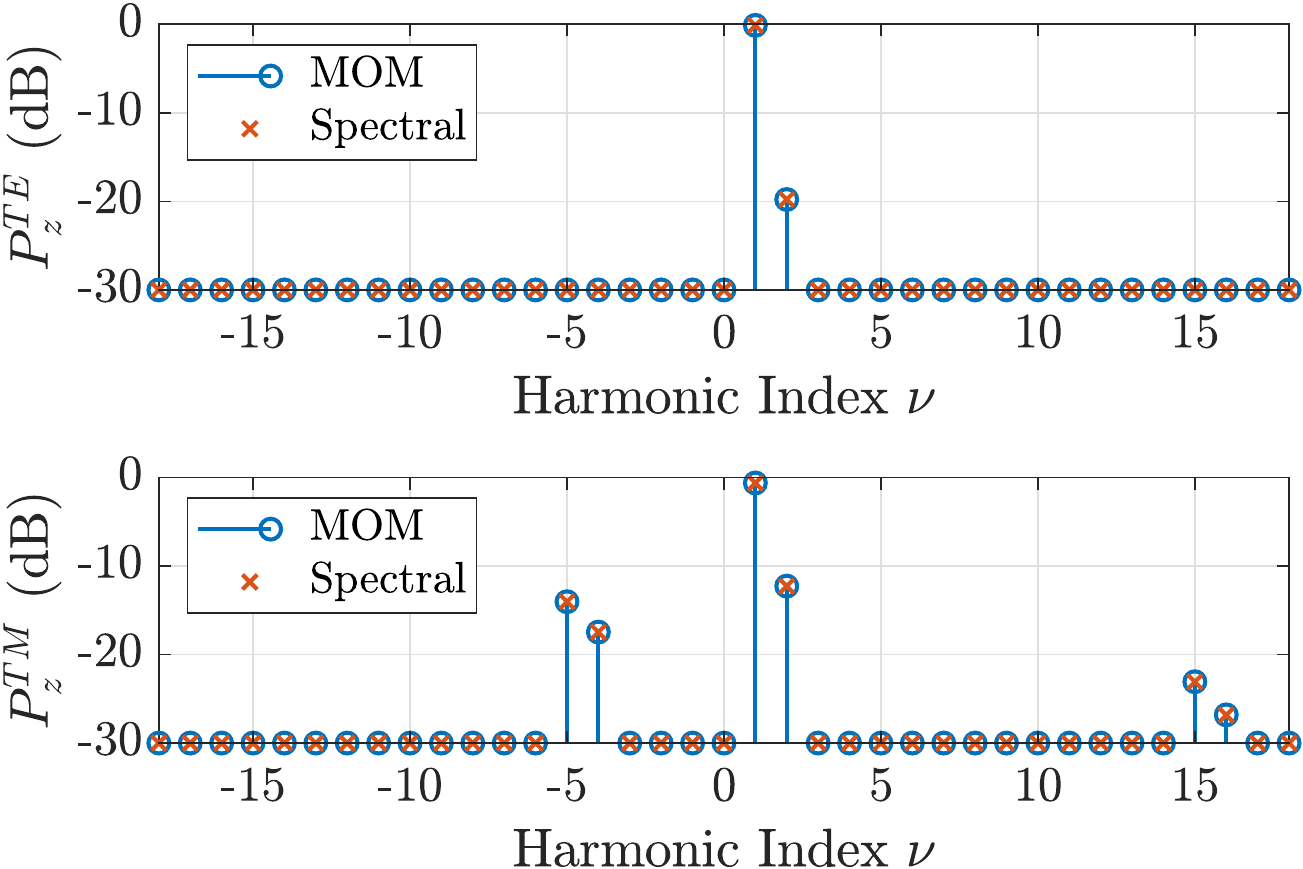}
	\caption{Normal power radiated at each frequency $f_0+\nu f_s$ from the metasurface in Fig. \ref{fig:cap_strips} modulated as specified in case A of Table \ref{tab:params}. In this case, the metasurface up-converts and deflects the scattered field. It can be seen that the MoM solver and spectral-domain methods are in good agreement.}
	\label{fig:pow_20p}
	\end{figure}

The spectrum computed by the spectral-domain method as well as the finest MoM solution is shown in Fig. \ref{fig:pow_20p}. As expected, the sawtooth wave up-converts the scattered wave to $f_0 + f_s$ and deflects it to $42.3^{\circ}$. The expansion in (\ref{eq:e_exp}) used by the spectral-domain method was truncated at $p'=\pm40$ and $\nu=\pm720$ for TE illumination and at $p'=\pm100$ and $\nu=\pm49$ for TM illumination. The error energy between the spectral-domain method and the MoM solver is $\Delta(\bar E^{\rm{(MoM)}},\bar E^{\rm{(Spec)}})=4.2\times 10^{-3}$ for the TE$_z$ simulation and $\Delta(\bar E^{\rm{(MoM)}},\bar E^{\rm{(Spec)}})=5.0\times 10^{-3}$ for the TM$_z$ simulation, confirming the validity of the MoM solver.

\subsection{Subwavelength Modulation Period}\label{sec:results_subharmmix}
	\begin{figure}
	\centering
	\includegraphics[width=\columnwidth]{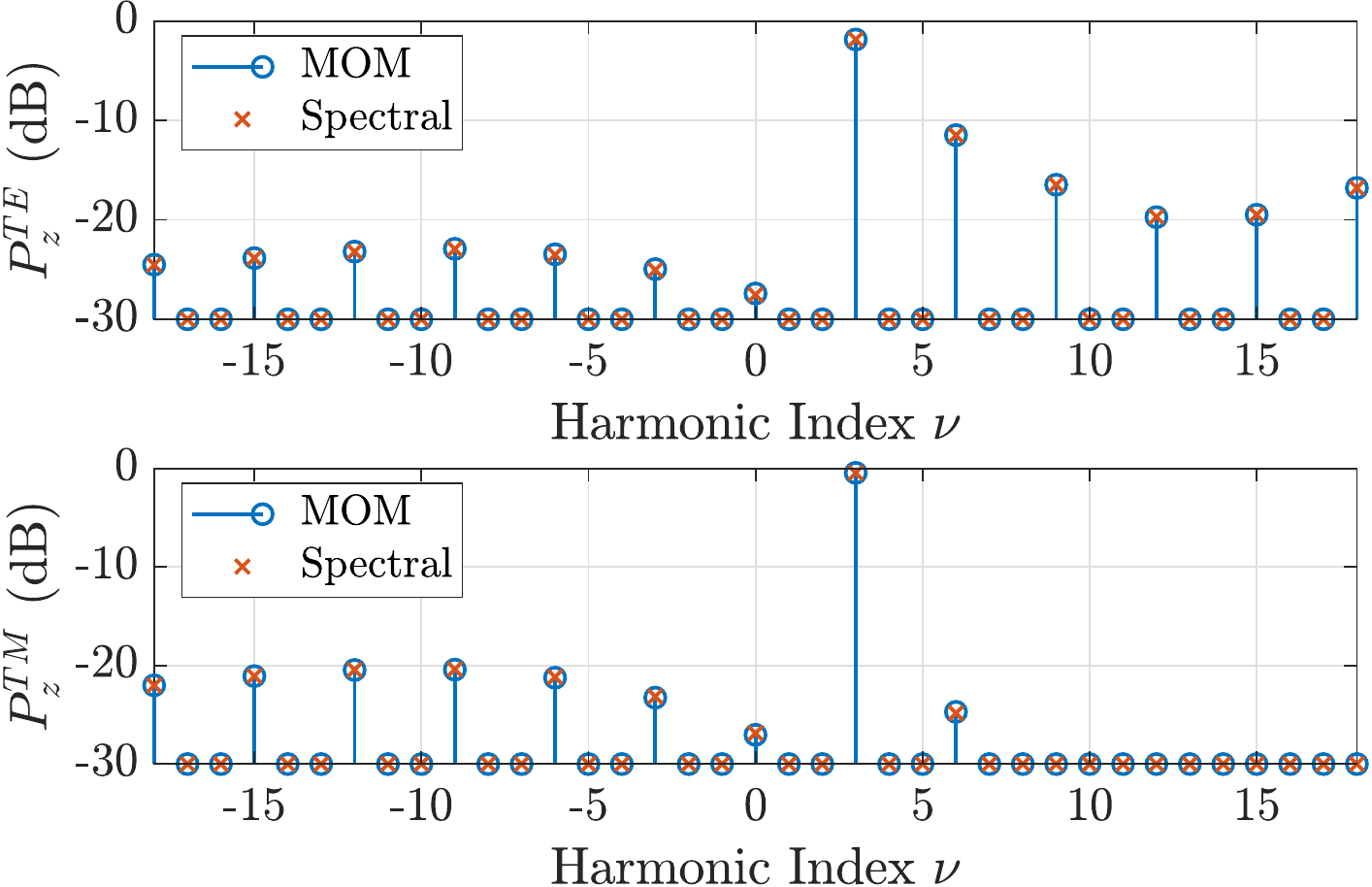}
	\caption{Normal power radiated at each frequency $f_0+\nu f_s$ from the metasurface in Fig. \ref{fig:cap_strips} modulated as specified in case B of Table \ref{tab:params}. In this case, the metasurface up-converts the scattered field to a higher-order harmonic of the modulation (sub-harmonic mixing). This behavior does not arise in the continuous model of traveling-wave modulation.}
	\label{fig:pow_3p}
	\end{figure}
	
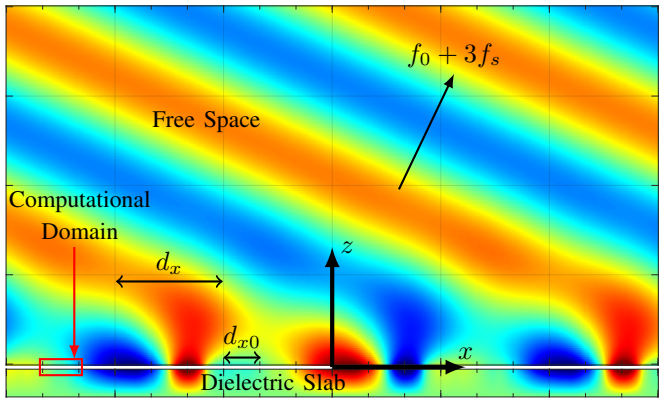
\begin{figure}
\centering
\begin{tikzpicture}
    \node[anchor=south west,inner sep=0](image)at(0,0,0){\includegraphics[clip,trim=0cm 0cm 0cm 0cm,width=\columnwidth]{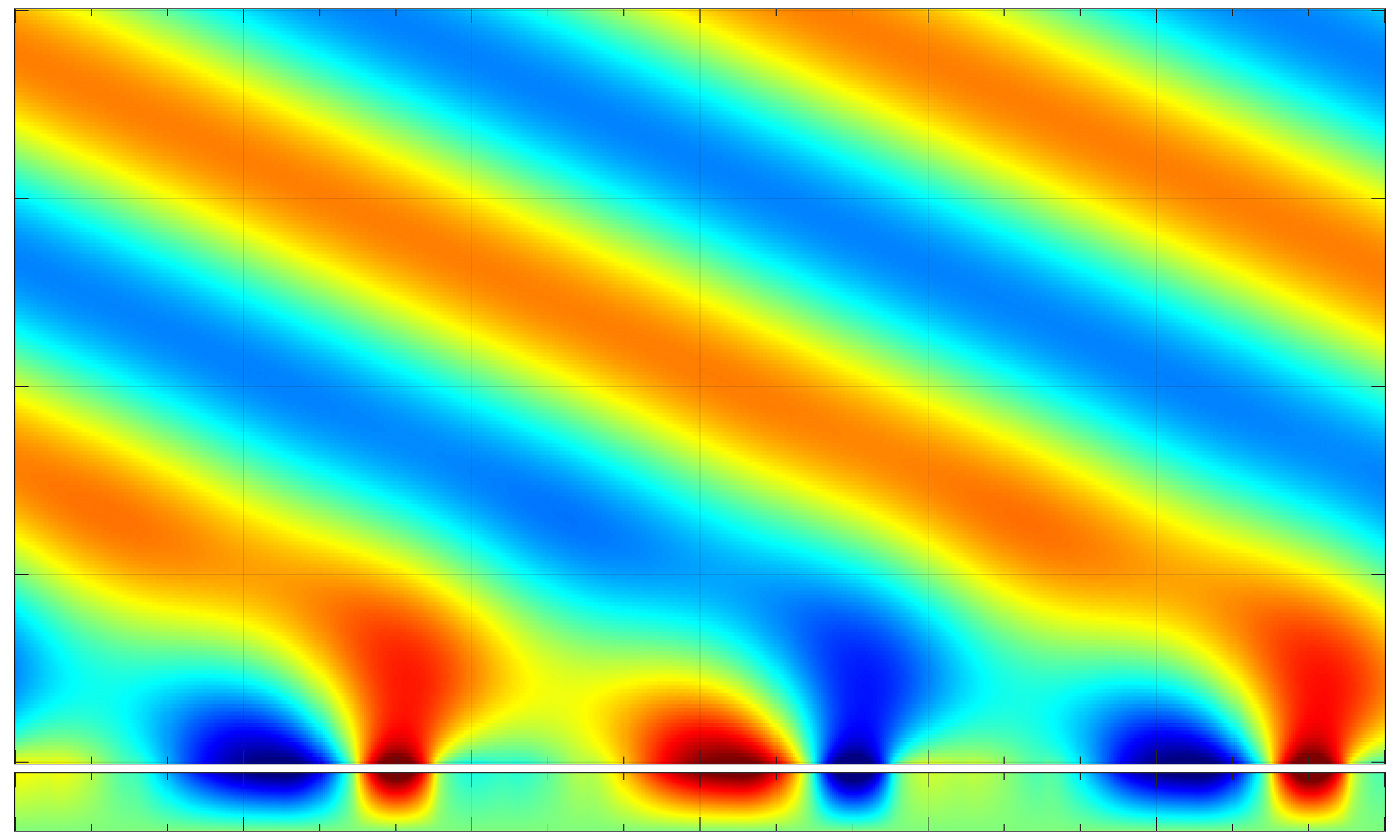}};
	\definecolor{myred}{rgb}{0.40,0.00,0.60};
    \begin{scope}[x={(image.south east)},y={(image.north west)},font=\fontsize{10}{6}]
        \begin{scope}[shift={(0.5,0.086)}] 
          \draw[black,ultra thick,-latex] coordinate (o) (0,0) -- coordinate (a) (0:.2) node[above=-1pt] {$x$};
          \draw[black,ultra thick,-latex] (o) -- coordinate (c)(90:.3) node[right=-1pt] {$z$};
        \end{scope}
        \begin{scope}[shift={(0.365,0.11)}] 
          \draw[black,thick,<->] (-\dx,0) -- (\dx,0) node[midway,above=0pt] {$d_{x0}$};
        \end{scope}
        \begin{scope}[shift={(0.256,0.3)}] 
          \draw[black,thick,<->] (-\bigdx,0) -- (\bigdx,0) node[midway,above=-1pt] {$d_{x}$};
        \end{scope}
        \begin{scope}[shift={(0.6,0.53)}] 
          \draw[black,thick,-latex] coordinate (o) (0,0) -- coordinate (a) (74:.3) node[above=-1pt] {$f_0 + 3 f_s$};
        \end{scope}
        \begin{scope}[shift={(0.062,0.086)}] 
          \draw[red,thick] (0,-\dy) rectangle (.063,\dy);
          \draw[red,thick,-latex] (0.5*.063+.02,.3) -- (0.5*.063+.02,\dy);
			\node[align=center,text width=2cm, text=black,rotate=0] (c) at (.06,.3+.08) {{\small Computational Domain}};
        \end{scope}
        \begin{scope}[font=\fontsize{8}{6}]
			\node[align=left,text width=3cm, text=black,rotate=0] (c) at (.472,0.048) {{\small Dielectric Slab}};
			\node[align=left,text width=3cm, text=black,rotate=0] (c) at (.4,0.7) {{\small Free Space}};
		\end{scope}
		
    \end{scope}
\end{tikzpicture}
\caption{Time snapshot of the scattered electric field profile for the metasurface in Fig. \ref{fig:cap_strips} modulated as described in case B of Table \ref{tab:params} for TE$_z$ incidence ($\bar E = E_y \hat y$). Note that the two regions are not drawn to scale. The scattered plane wave at the up-converted frequency $f_0 + 3 f_s$ can be seen propagating away from the metasurface into free space.}
\label{fig:spatial_field}
\end{figure}

In this example, results of the MoM and spectral-domain solvers will be compared for the metasurface modulation described in case B of Table \ref{tab:params}. The only difference between cases A and B is that the number of stixels per supercell is reduced from 20 to 3. Since the stixel size remains the same, this reduces the supercell size from $4\lambda_0$ to $0.6\lambda_0$. As a result, propagation is only supported for frequencies $f_0 + p\times 3 f_s$ for $p \in \mathbb{Z}$. Thus, when a sawtooth phase modulation waveform is applied to the metasurface, we expect the reflected signal to be up-converted from $f_0$ to $f_0 + 3 f_s$ \cite{zca}. 
This result is confirmed in Fig. \ref{fig:pow_3p}. A snapshot of the spatial profile of the scattered electric field for the TE$_z$ case is shown in Fig. \ref{fig:spatial_field}. Since the modulation frequency is small with respect to the RF carrier frequency ($f_s / f_0 = 2.5 \ \times 10^{-6}$), all the reflected energy is directed in the specular direction. In the MoM simulation, the current expansion included $M=301$ spatial samples and $U=601$ frequency harmonics. The expansion in (\ref{eq:e_exp}) used by the spectral-domain method was truncated at $p'=\pm50$ and $\nu=\pm300$. The error energy between the MoM and spectral-domain solvers is $\Delta(\bar E^{\rm{(MoM)}},\bar E^{\rm{(Spec)}})=9.9\times 10^{-4}$ for TE$_z$ polarization and $\Delta(\bar E^{\rm{(MoM)}},\bar E^{\rm{(Spec)}})=8.7 \times 10^{-3}$ for TM$_z$ polarization.

\FPset\dx{.028}
\FPset\dy{.02}
\FPset\bigdx{.08}

\subsection{Continuum Limit - Sinusoidal Traveling-Wave}\label{sec:results_sin}

\begin{figure}
	\centering
	\includegraphics[width=\columnwidth]{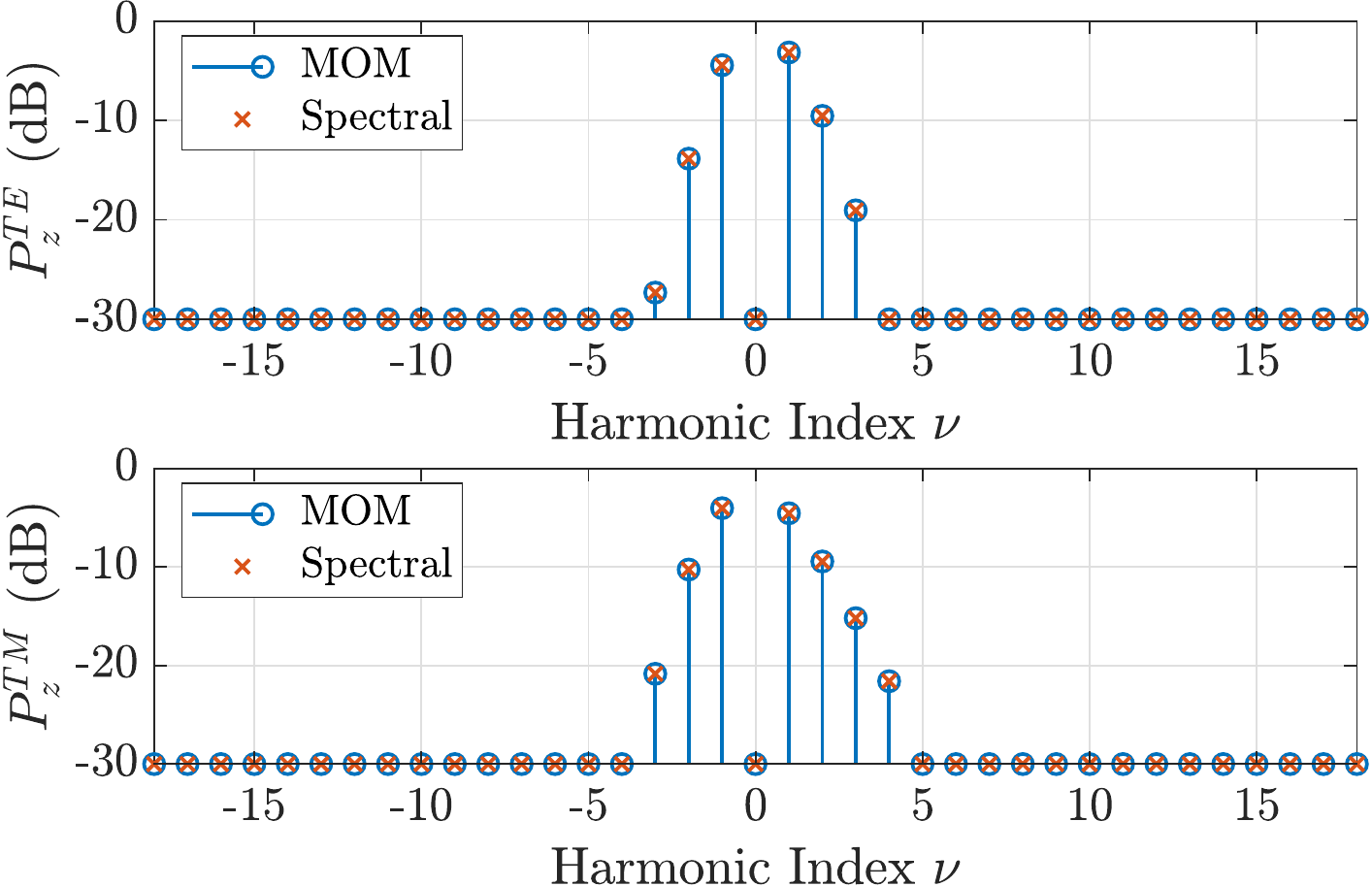}
	\caption{Normal power radiated at each frequency $f_0+\nu f_s$ from the metasurface in Fig. \ref{fig:cap_strips} modulated as specified in case C of Table \ref{tab:params}. Since this example is approaching continuous modulation, each frequency propagates in a unique direction. As a result, this modulation scheme depletes the fundamental frequency $f_0$ as well as the specular reflection from the scattered field. }
	\label{fig:pow_sin}
	\end{figure}

As a final example, we consider a nearly continuous structure described by case C of Table \ref{tab:params}. In this case, there are 60 stixels included in a spatial period, and the stixel size is brought down to $\lambda_0 / 10$ (thus the supercell size is $6\lambda_0$). Additionally, we assume that the sheet capacitance is sinusoidally modulated at a frequency of 500 MHz. For a spatially uniform sheet capacitance, we observe from (\ref{eq:phi}) that the $\nu^{\rm{th}}$ Fourier series coefficient of the reflection coefficient, $\Gamma_{\nu}$, can be computed as
\begin{equation} \label{eq:•}
{\Gamma _\nu } = \frac{1}{{{T_s}}}\int\limits_{t = 0}^{{T_s}} {{e^{ - 2j{{\tan }^{ - 1}}\left( {Z_0^{{\rm{TX}}}{\omega _0}C_0^{{\rm{TX}}}{\Delta _C}\left( t \right)} \right)}}{e^{ - j\nu {\omega _s}t}}dt} .
\end{equation}
Thus, if ${Z_0^{{\rm{TX}}}{\omega _0}C_0^{{\rm{TX}}}{\Delta _C}\left( t \right)}=A\cos \omega_s t$, then the fundamental harmonic of the reflection coefficient is given by
\begin{equation} \label{eq:•}
{\Gamma _0} = \frac{1}{{{T_s}}}\int\limits_{t = 0}^{{T_s}} {{e^{ - 2j{{\tan }^{ - 1}}\left( {A\cos {\omega _s}t} \right)}}dt}  = \frac{2}{{\sqrt {1 + {A^2}} }} - 1 .
\end{equation}
When $A= \sqrt{3}$, the fundamental harmonic of the reflection coefficient (and thus the RF carrier frequency of the reflected field) goes to zero.
Now suppose we add a spatial dependence to the modulation function in the form of a continuous traveling-wave, i.e. $C(t) \rightarrow C(t-x T_s / d_x)$ where $d_x$ is the spatial modulation period.
In this case, when $A= \sqrt{3}$, there should not be any power radiated at the fundamental frequency nor in the specular direction. In other words, the reflected wave is completely redistributed in frequency and spatial spectrum.
From Fig. \ref{fig:pow_sin}, we observe that the reflected power at the RF carrier frequency is zero. Since $\nu=0$ corresponds to the only frequency which radiates in the specular direction, we can also conclude that no power is reflected back at the specular angle of $25^\circ$. In the MoM simulation, the current expansion included $M=301$ spatial samples and $U=101$ frequency harmonics. The expansion in (\ref{eq:e_exp}) used by the spectral-domain method was truncated at $p'=\pm75$ and $\nu=\pm50$.
The error energy between the MoM and spectral-domain solvers is $\Delta(\bar E^{\rm{(MoM)}},\bar E^{\rm{(Spec)}})=4.4 \times 10^{-6}$ for TE$_z$ polarization and $\Delta(\bar E^{\rm{(MoM)}},\bar E^{\rm{(Spec)}})=4.0 \times 10^{-4}$ for TM$_z$ polarization.

\section{Conclusion} \label{sec:conclusion}
As research into traveling-wave modulated structures progresses, it is vital to develop accurate computational methods capable of efficiently simulating practical designs.
Typically, spatially-discrete traveling-wave modulation (SD-TWM) is employed in the physical implementation of traveling-wave modulated structures.
In each spatial period (i.e. supercell) of a SD-TWM structure, a staggered modulation signal is applied to a discrete array of sub-cells (i.e. stixels).
Unlike the continuous limit, where closed form solutions are often available, numerical methods must be used to analyze SD-TWM.
Due to the complicated space-time dependence of SD-TWM structures, fine discretization is required to obtain an adequate model.
Further, the computational methods reported to date require the computational domain to extend in space over an entire supercell.
As a result, the computational cost can become prohibitive; particularly for simulating complex patterned structures or nearly continuous structures with a large number of stixels per supercell.
However, by taking advantage of the space-time symmetry of SD-TWM structures, we have shown that the fields within a single stixel determine the entire problem.
Therefore, the computational domain can be reduced from an entire supercell to a single stixel.
The simplicity of adding this symmetry to computational electromagnetic solvers will enable researchers to study SD-TWM for a wide variety of designs.

In this paper, a relation between neighboring stixels of a SD-TWM structure was derived and incorporated into a method of moments analysis. 
The derived boundary condition, referred to as the interpath relation, can be interpreted as a modified frequency-harmonic-dependent periodic boundary condition.
Using the interpath relation, the fields within a single stixel are sufficient to solve the entire domain. 
It was shown that the interpath relation can be incorporated into a method of moments solver by slightly modifying the basis functions used to expand the surface current.
As a result the number of unknowns required to solve the problem are reduced by a factor of $L$, the number of stixels in a supercell.
The method was applied to a SD-TWM sheet capacitance over a grounded dielectric and benchmarked against a spectral-domain solver.
Various SD-TWM examples were explored to both motivate and validate the proposed method.

\appendices
\section{Derivation of the Spectral-Domain Dyadic Green's Function for a Grounded Dielectric}\label{sec:A1}

\FPset\lenax{0.2}
\FPset\lenay{0.185}
\FPset\lenaz{0.25}
\FPset\lent{.01}
\FPset\rx{-35}
\FPset\ry{32}
\begin{figure}[!t]
\centering
\begin{tikzpicture}
	\definecolor{myred}{rgb}{0.8,0.1,0.1};
	\definecolor{mypurp}{rgb}{0.7,0.1,0.8};
	\definecolor{myblue}{rgb}{0,.4,1};
	\definecolor{mygreen}{rgb}{0,.7,.4};
    \node[anchor=south west,inner sep=0](image)at(0,0,0){\includegraphics[clip,trim=1cm 4cm 0cm 3cm,width=\columnwidth]{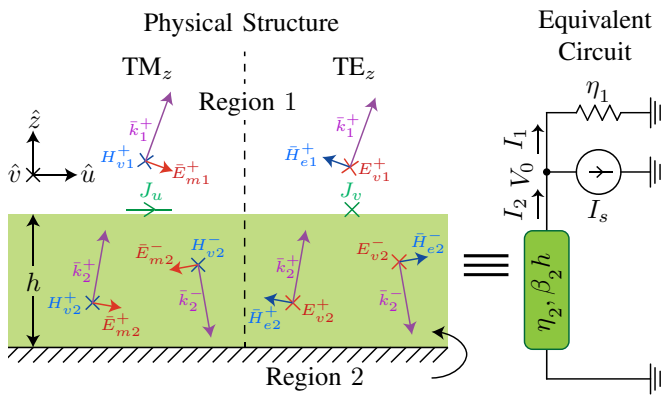}};
    \begin{scope}[x={(image.south east)},y={(image.north west)},font=\fontsize{10}{6}]
        
		\node[align=center,rotate=0] (c) at (0.35,1.07) {Physical Structure};
		\node[align=center,rotate=0] (c) at (0.88,1.04) {Equivalent \\ Circuit};		
		
		\begin{scope}[shift={(.095,-.095)}]
			\node[align=center,rotate=90] (c) at (0.675,0.75) {$V_0$};
			\node[align=center,rotate=0] (c) at (0.79,0.65) {$I_s$};
			\node[align=center,rotate=90] (c) at (0.665,0.655) {$I_2$};
			\node[align=center,rotate=90] (c) at (0.665,0.85) {$I_1$};
			\node[align=center,rotate=90] (c) at (0.712,0.425){$\eta_{\mm{-3}{2}},\beta_{\mm{-3}{2}}h$};
			\node[align=center,rotate=0] (c) at (0.79,0.98) {$\eta_{\mm{-2}{1}}$};
		\end{scope}
		
		\begin{scope}[font=\fontsize{8}{6}]
			\node[mygreen,align=center,rotate=0] (c) at (0.215,0.61) {$J_u$};
			\node[mygreen,align=center,rotate=0] (c) at (0.515,0.61) {$J_v$};
		\end{scope}

		\begin{scope}[font=\fontsize{6}{6}]
			\node[myred,align=center,rotate=0](c)at(0.275,0.66){$\bar E_{m1}^+$};
			\node[myblue,align=center,rotate=0](c)at(0.165,0.705){$H_{v1}^+$};
			\node[mypurp,align=center,rotate=0](c)at(0.2,0.78){$\bar k_1^+$};
			\node[myred,align=center,rotate=0](c)at(0.55,0.67){$E_{v1}^+$};
			\node[myblue,align=center,rotate=0](c)at(0.44,0.71){$\bar H_{e1}^+$};
			\node[mypurp,align=center,rotate=0](c)at(0.51,0.79){$\bar k_1^+$};
		\end{scope}
		\begin{scope}[shift={(0.0,0.0)},font=\fontsize{6}{6}]
			\node[myred,align=center,rotate=0](c)at(0.17,0.24){$\bar E_{m2}^+$};
			\node[myblue,align=center,rotate=0](c)at(0.085,0.3){$H_{v2}^+$};
			\node[mypurp,align=center,rotate=0](c)at(0.115,0.39){$\bar k_2^+$};
			\node[myred,align=center,rotate=0](c)at(0.22,0.44){$\bar E_{m2}^-$};
			\node[myblue,align=center,rotate=0](c)at(0.30,0.46){$H_{v2}^-$};
			\node[mypurp,align=center,rotate=0](c)at(0.275,0.31){$\bar k_2^-$};
			\node[myred,align=center,rotate=0](c)at(0.465,0.28){$E_{v2}^+$};
			\node[myblue,align=center,rotate=0](c)at(0.385,0.26){$\bar H_{e2}^+$};
			\node[mypurp,align=center,rotate=0](c)at(0.42,0.4){$\bar k_2^+$};
			\node[myred,align=center,rotate=0](c)at(0.547,0.45){$E_{v2}^-$};
			\node[myblue,align=center,rotate=0](c)at(0.63,0.47){$\bar H_{e2}^-$};
			\node[mypurp,align=center,rotate=0](c)at(0.575,0.31){$\bar k_2^-$};
		\end{scope}
		\node[black,align=center,rotate=0](c)at(0.21,0.95){ TM$_z$};
		\node[black,align=center,rotate=0](c)at(0.52,0.95){TE$_z$};
		\node[black,align=center,rotate=0](c)at(0.46,0.09){Region $2$};
		\node[black,align=center,rotate=0](c)at(0.36,0.85){Region $1$};
		
		\node[black,align=center,rotate=0](c)at(0.038,0.36){$h$};
		\node[black,align=center,rotate=0](c)at(0.04,0.81){$\hat z$};
		\node[black,align=center,rotate=0](c)at(0.12,0.66){$\hat u$};
		\node[black,align=center,rotate=0](c)at(0.01,0.66){$\hat v$};
		
    \end{scope}
\end{tikzpicture}
\caption{An illustration of a grounded dielectric and its equivalent circuit model when excited by planar current sheets of the form ${\bar J_0}{e^{ - j\left( {{k_x}x + {k_y}y} \right)}}$. The $\hat u$ axis is defined to be along the direction of $k_x\hat x + k_y \hat y$. Thus, the $\hat u$ component of $\bar J_0$ excites TM$_z$ waves, while the $\hat v$ component of $\bar J_0$ excites TE$_z$ waves.}
\label{fig:greensFcn}
\end{figure}

To obtain the dyadic Green's function used in the MoM formulation, we will impress current sheets of the form ${\bar J_0}{e^{ - j\left( {{k_x}x + {k_y}y} \right)}}$ on the surface of the grounded dielectric as shown in Fig. \ref{fig:greensFcn}.
We will define region 1 to be free space and region 2 to be the dielectric substrate. As discussed in Section \ref{sec:mom}, the substrate has a thickness $h$ and a dielectric constant of $\epsilon_r$. Further, we define ${{\bar k}_\rho } = {k_x}\hat x + {k_y}\hat y$. This allows us to define two orthogonal vectors $\hat u$ and $\hat v$ such that TM$_z$ and TE$_z$ excitations can be handled independently. The unit vector $\hat v$ is defined to be orthogonal to $\bar k_\rho$, and is given by
\begin{equation} \label{eq:v}
\hat v = \frac{{\hat z \times {{\bar k}_\rho }}}{{\left| {\hat z \times {{\bar k}_\rho }} \right|}} .
\end{equation}
The unit vector $\hat u$ is defined to be orthogonal to $\hat v$ and $\hat z$ (i.e. $\hat u$ is the unit vector in the direction of $\bar k _\rho $), and is given by
\begin{equation} \label{eq:u}
\hat u = \hat v \times \hat z = \frac{\bar k _\rho}{\left| \bar k _\rho \right|} .
\end{equation}
We will decompose $\bar J_0$ into two orthogonal components. The component $J_u=\bar J_0 \cdot \hat u$ excites TM$_z$ waves, while $J_v=\bar J_0 \cdot \hat v$ excites TE$_z$ waves. Since $E_u$ depends only on $J_u$ and $E_v$ depends only on $J_v$, the Green's function will be a diagonal tensor in the $u$-$v$ coordinate system. The Green's function is defined such that \cite{jin}
\begin{equation} \label{eq:Gnorm}
{{\bar E}_{0t}} =  - j{k_0}{Z_0}\tensor{G}\left(k_x,k_y\right){{\bar J}_0} ,
\end{equation}
where ${{\bar E}_{0t}}{e^{ - j\left( {{k_x}x + {k_y}y} \right)}}$ is the component of the electric field produced by the impressed current in the $x$-$y$ plane.

The fields excited in both regions can be decomposed into plane waves. In region 1, only upward-traveling waves are excited since the medium is unbounded from above. Meanwhile, in region 2, both upward- and downward-traveling waves are excited by the current sheets. When $J_u$ is non-zero, TM$_z$  waves are excited in regions 1 and 2. As shown in Fig. \ref{fig:greensFcn}, the magnetic field for this excitation contains a single component along the $\hat v$ direction; while the electric field contains components along both the $\hat u$ and $\hat z$ directions. Conversely, when $J_v$ is non-zero, TE$_z$ waves are excited in regions 1 and 2. As shown in Fig. \ref{fig:greensFcn}, the electric field for this excitation contains a single component along the $\hat v$ direction; while the magnetic field contains components along both the $\hat u$ and $\hat z$ directions.

For both the TM$_z$ and TE$_z$ cases, a transmission line model can be used to find the transverse component of electric field at the interface between regions 1 and 2 \cite{Lecture}. 
As shown in Fig \ref{fig:greensFcn}, the transmission line model contains a shorted transmission line representing region 2, a current source representing the impressed current at the interface, and a load representing region 1. Using this model, computing the transverse component of electric field at the interface simply becomes a matter of finding $V_0$ in the equivalent circuit shown in Fig. \ref{fig:greensFcn}. For TM$_z$ excitations, we set $I_s=J_u$ and make the substitutions provided in the first column of Table \ref{tab:substitutions}. The transverse component of electric field in the $\hat u$ direction is found by computing $V_0$. For TE$_z$ excitations, we set $I_s=J_v$ and make the substitutions provided in the second column of Table \ref{tab:substitutions}. The transverse component of electric field in the $\hat v$ direction is found by computing $V_0$.

\renewcommand{\arraystretch}{2}
\begin{table}[]
\centering
\caption{TM$_z$ and TE$_z$ Substitutions}
\label{tab:substitutions}
\begin{tabular}{|c|c|}
\hline
\textbf{TM$_z$} & \textbf{TE$_z$} \\ \hline

$I_s=J_u$ & $I_s=J_v$ \\ \hline

${\beta _1} = \sqrt {k_0^2 - {{\left| {{{\bar k}_\rho }} \right|}^2}} $
& ${\beta _1} = \sqrt {k_0^2 - {{\left| {{{\bar k}_\rho }} \right|}^2}} $ \\ \hline

${\eta _{\mm{-2}{1}}} = \frac{{{\beta _1}}}{{\omega {\epsilon _0}}}  = {Z_0}\frac{{{\beta _1}}}{{{k_0}}}$
& ${\eta _{\mm{-2}{1}}} = \frac{{\omega {\mu _0}}}{{{\beta _1}}}  = {Z_0}\frac{{{k_0}}}{{{\beta _1}}} $ \\ \hline

${\beta _2} = \sqrt {{\epsilon _r}k_0^2 - {{\left| {{{\bar k}_\rho }} \right|}^2}} $ 
& ${\beta _2} = \sqrt {{\epsilon _r}k_0^2 - {{\left| {{{\bar k}_\rho }} \right|}^2}} $ \\ \hline

${\eta _{\mm{-2}{2}}} = \frac{{{\beta _2}}}{{\omega {\epsilon _r}{\epsilon _0}}} = {Z_0}\frac{{{\beta _2}}}{{{\epsilon _r}{k_0}}}$
& ${\eta _{\mm{-2}{2}}} = \frac{{\omega {\mu _0}}}{{{\beta _2}}} = {Z_0}\frac{{{k_0}}}{{{\beta _2}}}$ \\ \hline

${\hat u \cdot \bar E_{0t}} = {V_0}$
& ${\hat v \cdot \bar E_{0t}} = {V_0}$ \\ \hline
\end{tabular}
\end{table}

It is clear that, for either TM$_z$ or TE$_z$ excitations, we must solve for $V_0$ in the equivalent circuit shown in Fig. \ref{fig:greensFcn}. The impedance looking into the shorted transmission line is given by $ Z_2 = j{\eta _2}\tan {\beta _2}h$. Therefore, the total impedance, $Z_t$, seen by the source is a parallel combination of $\eta_1$ and $Z_2$, given by
\begin{equation} \label{eq:Zt}
{Z_t} = \frac{Z_0}{{\frac{Z_0}{{{\eta _1}}} - j\frac{Z_0}{{{\eta _2}}}\cot \beta_2 h}} .
\end{equation}
Substituting the expressions for $\eta_1$ and $\eta_2$ from Table \ref{tab:substitutions} into (\ref{eq:Zt}), we obtain
\begin{equation} \label{eq:ZtTM}
Z_t^{{\rm{TM}}_z} = \frac{{{Z_0}}}{{\frac{{{k_0}}}{{{\beta _1}}} - j\frac{{{\varepsilon _r}{k_0}}}{{{\beta _2}}}\cot \beta_2 h}} 
\end{equation}
\begin{equation} \label{eq:ZtTE}
Z_t^{{\rm{TE}}_z} = \frac{{{Z_0}}}{{\frac{{{\beta _1}}}{{{k_0}}} - j\frac{{{\beta_2 }}}{{{k_0}}}\cot \beta_2 h}} .
\end{equation}
Since $V_0=-Z_t I_s$ (the minus sign is due to the orientation of $I_s$), the $u$-$v$ components of $\bar E_{0t}$ can be computed as
\begin{equation} \label{eq:Eu}
\hat u \cdot {{\bar E}_{0t}} = - j{k_0}{Z_0} G_u J_u 
\end{equation}
\begin{equation} \label{eq:Ev}
\hat v \cdot {{\bar E}_{0t}} = - j{k_0}{Z_0} G_v J_v ,
\end{equation}
where
\begin{equation} \label{eq:Gu}
G_u = \frac{{1/j{k_0}}}{{\frac{{{k_0}}}{{{\beta _1}}} - j\frac{{{\varepsilon _r}{k_0}}}{{{\beta _2}}}\cot {\beta _2}h}} 
\end{equation}
\begin{equation} \label{eq:Gv}
G_v = \frac{{1/j{k_0}}}{{\frac{{{\beta _1}}}{{{k_0}}} - j\frac{{{\beta _2}}}{{{k_0}}}\cot \beta_2 h}} .
\end{equation}
In these expressions, $G_u$ and $G_v$ represent the dyadic Green's function in the $u$-$v$ coordinate system. To obtain $\tensor{G}$ in the $x$-$y$ coordinate system, we first define $\psi$ such that $\tan \psi = k_y/k_x$. We can subsequently write $\tensor{G}$ as
\renewcommand{\arraystretch}{1}
\begin{equation} \label{eq:Grot}
\tensor{G}\left( k_x , k_y \right) = R\left(\psi\right)
\left[ {\begin{array}{*{20}{c}}
{{G_u}}&0\\
0&{{G_v}}
\end{array}} \right]
R\left(-\psi\right) ,
\end{equation}
where $R\left(\psi\right)$ is the rotation matrix given by
\begin{equation} \label{eq:R}
R\left(\psi\right) = \left[ {\begin{array}{*{20}{c}}
{\cos \psi }&{- \sin \psi }\\
{ \sin \psi }&{\cos \psi }
\end{array}} \right] .
\end{equation}
Carrying out the matrix multiplication yields the following expressions for the elements of $\tensor{G}\left(k_x, k_y \right)$ :
\begin{equation} \label{eq:Gxx}
\hat x \cdot \tensor{G}\left(k_x, k_y \right) \cdot \hat x = {G_u}{\cos ^2}\psi  + {G_v}{\sin ^2}\psi 
\end{equation}
\begin{equation} \label{eq:Gyy}
\hat y \cdot \tensor{G}\left(k_x, k_y \right) \cdot \hat y = {G_u}{\sin ^2}\psi  + {G_v}{\cos ^2}\psi 
\end{equation}
\begin{align} \label{eq:Gxy}
\begin{split}
\hat x \cdot \tensor{G}\left(k_x, k_y \right) \cdot \hat y & = 
\hat y \cdot \tensor{G}\left(k_x, k_y \right) \cdot \hat x \\ 
& =\cos \psi \sin \psi \left( {{G_u} - {G_v}} \right)
\end{split} .
\end{align}
Thus the spectral-domain dyadic Green's function has been derived.

\section*{Acknowledgment}

This work was supported under the AFOSR MURI program FA9550-18-1-0379.

This research was supported in part by the computational resources and services provided by Advanced Research Computing at the University of Michgan in Ann Arbor, Michigan.

\ifCLASSOPTIONcaptionsoff
  \newpage
\fi

\bibliographystyle{IEEEtran}
\bibliography{references}

\begin{IEEEbiography}
[{\includegraphics[width=.9in,clip,keepaspectratio]{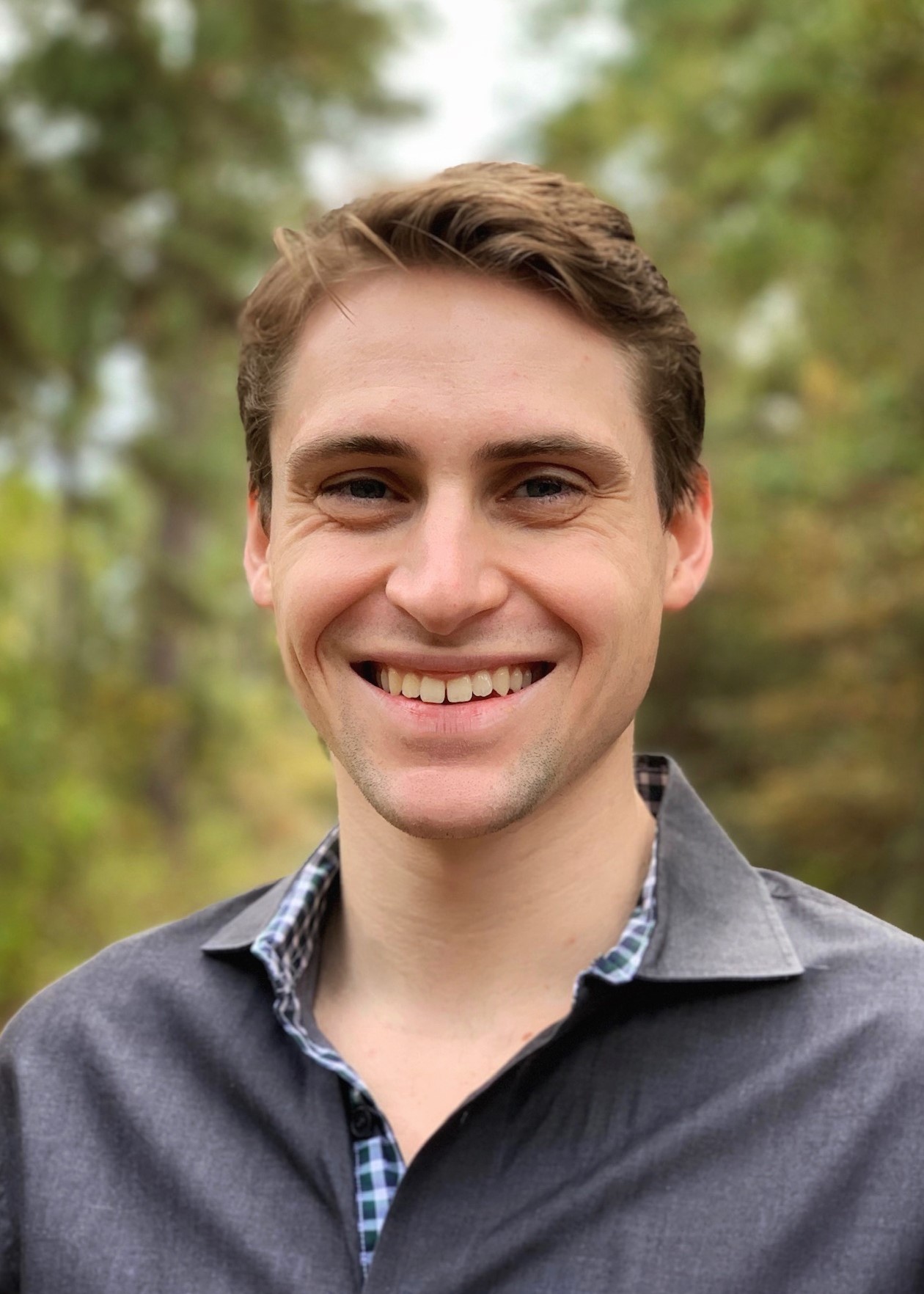}}]{Cody Scarborough}
graduated from the University of Texas at Austin with a B.S. in Electrical and Computer Engineering in 2017. Throughout his undergraduate degree, he was employed at Applied Research Laboratories and subsequently at the University of Texas as an undergraduate research assistant. His work has included projects in global navigation systems, wireless communication networks for vehicular technology and the efficient computation of radiation from cyrptographic integrated circuits. He is currently a graduate student pursuing a PhD at the University of Michigan, where he studies time-varying electromagnetic structures and metamaterials. His research interests include the study of wave phenomena, non-linear electronic devices, solid-state physics, spatio-temporal modulation, microwave systems and optics.
\end{IEEEbiography}

\begin{IEEEbiography}
[{\includegraphics[width=.9in,clip,keepaspectratio]{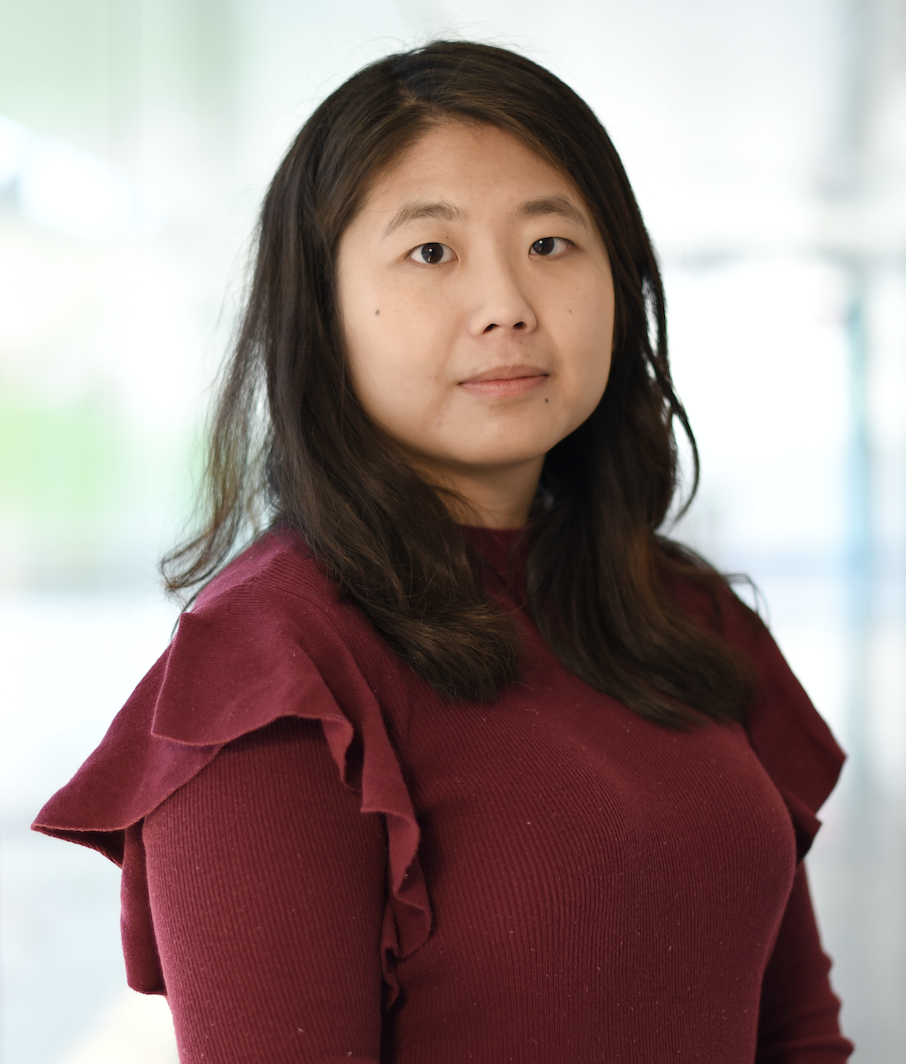}}]{Zhanni Wu}
received the B.S. degree from the School of Information Science and Engineering, Southeast University, Nanjing, China, in 2015, and the M.S.E. degree from the University of Michigan, Ann Arbor, MI, USA, in 2017, where she is currently pursuing the Ph.D. degree in electrical engineering. She was with the State Key Laboratory of Millimeter Waves, Southeast University, from 2013 to 2015, as an Undergraduate Research Assistant. Her current research interests include metamaterials/metasurfaces,
tunable metasurface devices, and wave propagation in spatio–temporal
modulated metamaterials.
\end{IEEEbiography}

\begin{IEEEbiography}[{\includegraphics[width=.9in ,clip,keepaspectratio]{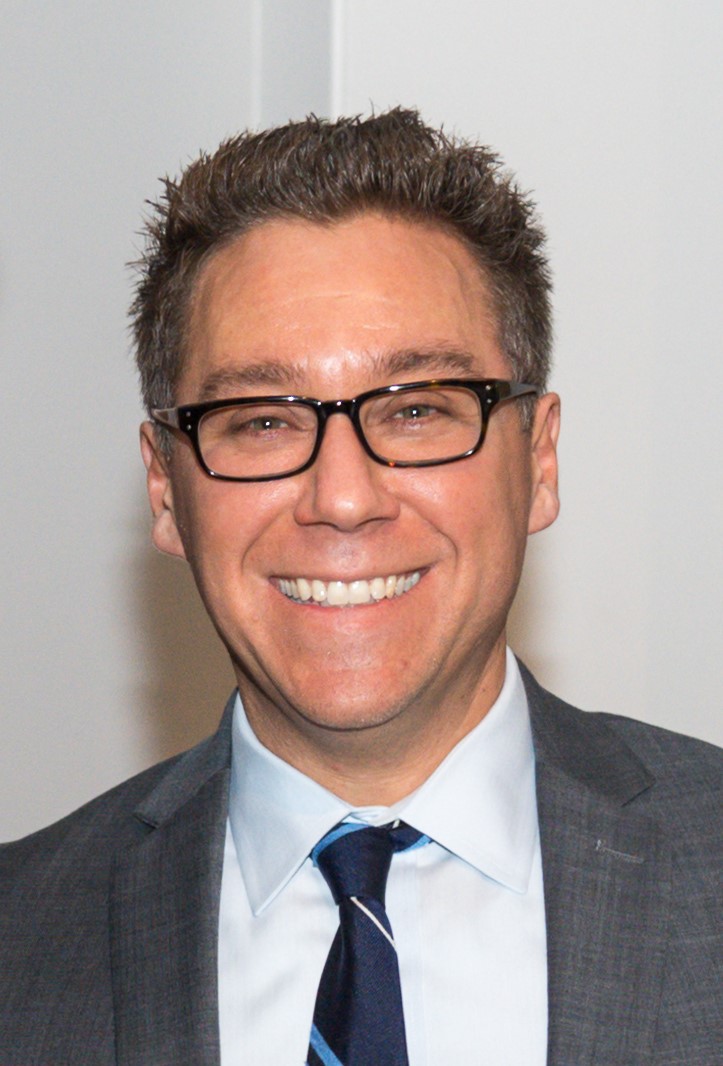}}]{Anthony Grbic}
received the B.A.Sc., M.A.Sc., and Ph.D. degrees in electrical engineering from the University of Toronto, Toronto, ON, Canada, in 1998, 2000, and 2005, respectively. In January 2006, he joined the Department of Electrical Engineering and Computer Science, University of Michigan, Ann Arbor, MI, USA, where he is currently a Professor. His research interests include engineered electromagnetic structures (metamaterials, metasurfaces, electromagnetic band-gap materials, frequency-selective surfaces), plasmonics, antennas, microwave circuits, wireless power transmission, and analytical electromagnetics/optics.

Dr. Grbic served as Technical Program Co-Chair in 2012 and Topic CoChair in 2016 and 2017 for the IEEE International Symposium on Antennas and Propagation and USNC-URSI National Radio Science Meeting. He was an Associate Editor for IEEE Antennas and Wireless Propagation Letters from 2010 to 2015. He is currently the Vice Chair of AP-S Technical Activities, Trident Chapter, IEEE Southeastern Michigan Section. Dr. Grbic was the recipient of AFOSR Young Investigator Award as well as NSF Faculty Early Career Development Award in 2008, the Presidential Early Career Award for Scientists and Engineers in January 2010. He also received an Outstanding Young Engineer Award from the IEEE Microwave Theory and Techniques Society, a Henry Russel Award from the University of Michigan, and a Booker Fellowship from the United States National Committee of the International Union of Radio Science in 2011. He was the inaugural recipient of the Ernest and Bettine Kuh Distinguished Faculty Scholar Award in the Department of Electrical and Computer Science, University of Michigan in 2012.
\end{IEEEbiography}

\end{document}